\documentclass[12pt,twoside]{article}
\usepackage{psfig}
\usepackage{latexsym}
\usepackage{amsmath}
\newcommand{\VolumeHeader}{}
\newcommand{\VolumeSerial}{LNS}

\newcommand{\ActivityName}{ {\normalsize {\it 
Summer School on Astroparticle Physics and Cosmology }}\\}
\newcommand{\ActivityDate}{ {\normalsize {\it
Trieste, 17 June -- 5 July 2002}}}

\newcommand{\be}{\begin{equation}}
\newcommand{\ee}{\end{equation}}
\newcommand{\bea}{\begin{eqnarray}}
\newcommand{\eea}{\end{eqnarray}}

\newcommand{\h}{\hat}
\newcommand{\htt}{h^{^{TT}}}

\newcommand{\al}{\alpha}
\newcommand{\bt}{\beta}
\newcommand{\m}{\mu}
\newcommand{\n}{\nu}
\newcommand{\Itf}{\hbox{$\hbox{${ I }$}\kern-.60em 
	\raise.4ex \hbox{$-$}$}}

\newcommand{\gzuab}{g^{^{\!\!\! 0} \alpha \beta}}
\newcommand{\gzdmn}{g^{^{\!\!\! 0}}_{\mu \nu}}
\newcommand{\gzumdn}{g^{^{\!\!\! 0} {\mu}}_{\ \nu}}
\newcommand{\Rzdmn}{R{\! \!\! \hbox{\raise .6ex 
	\hbox{$^{^0}$} } }_{\!\!\mu \nu}}
\newcommand{\Tzdmn}{T{\! \!\! \hbox{\raise .6ex 
	\hbox{$^{^0}$} } }_{\!\!\mu \nu}}

\newcommand{\LectureHeader}{Gravitational
Waves from Perturbed Black Holes and Neutron Stars}

\begin{document}
\pagestyle{myheadings}
\markboth{\LectureHeader}{\VolumeHeader}
\markright{\VolumeHeader}



\begin{titlepage}


\title{Gravitational Waves from Perturbed Black Holes and Relativistic Stars}

\author{Luciano Rezzolla\thanks{\tt rezzolla@sissa.it}
\\[1cm]
{\normalsize
{\it SISSA, International School for Advanced Studies, Trieste, Italy.}} 
\\
{\normalsize
{\it INFN, Sezione di Trieste, Italy.}}\\
\\\\
\\[10cm]
{\normalsize {\it Lectures given at the: }}
\\
\ActivityName 
\\
\ActivityDate 
\\[1cm]
{\small \VolumeSerial} 
}
\date{}
\maketitle
\thispagestyle{empty}
\end{titlepage}

\baselineskip=14pt
\newpage
\thispagestyle{empty}


\begin{abstract}
These lectures aim at providing an introduction to the properties of
gravitational waves and in particular to those gravitational waves that
are expected as a consequence of perturbations of black holes and neutron
stars. Imprinted in the gravitational radiation emitted by these objects
is, in fact, a wealth of physical information. In the case of black
holes, a detailed knowledge of the gravitational radiation emitted as a
response to perturbations will reveal us important details about their
mass and spin, but also about the fundamental properties of the event
horizon. In the case of neutron stars, on the other hand, this
information can provide a detailed map of their internal structure and
tell us about the equation of state of matter at very high density, thus
filling-in a gap in energies and densities that cannot be investigated by
experiments in terrestrial laboratories.
\end{abstract}

\vspace{6cm}




\newpage
\thispagestyle{empty}
\tableofcontents

\newpage
\setcounter{page}{1}

\section{A Brief Introduction to Gravitational Waves}
\label{intro_gws}

	Being prepared for a Summer School, these lecture notes are meant
principally as a first introduction to the rather vast and rapidly
growing area of research that deals with gravitational waves. As a
result, these notes will simply provide the basic concepts of this field
of research and briefly review the most important results. The
bibliography is not meant to be complete and the sparse references are
used mostly as pointers to review articles where the topics of these
lectures find a more detailed discussion and where a complete
presentation of the literature can be found. In what follows I will
assume that the reader is proficient in basic tensor algebra, is familiar
with the fundamental concepts in General Relativity and has already
encountered a black hole solution to Einstein equations.
	
	The lecture notes are organized in three main parts. In the first
one, I rapidly introduce the basic properties of gravitational waves by
deriving a wave solution of the Einstein equations and discussing the
sense of the transverse-traceless gauge. A brief discussion about the
generation of gravitational waves will also be made. In the second part,
I discuss in some detail the perturbation properties of a Schwarzschild
black hole and the gravitational radiation that is expected as a
result. Finally, the third part is devoted to presenting the basic
elements of the theory of stellar perturbations and how stellar
perturbations can be used to generate gravitational waves. The discussion
will be restricted to spherical oscillations in the case of relativistic
stars and to nonradial oscillations in the case of Newtonian stars. A
brief classification of the modes of oscillations will also be presented
together with an introduction to the onset of non-axisymmetric
instabilities.

	These lectures notes are taken in part from the introductory
course on General Relativity given to the graduate students at SISSA and
is inspired by the corresponding discussions in a number of textbooks and
in particular by those in~\cite{mtw74,s84,di90}. A space-like signature
$(-,+,+,+)$ will be used, with Greek indices taken to run from 0 to 3 and
Latin indices from 1 to 3. Covariant derivatives are denoted with a
semi-colon and partial derivatives with a comma. Tensors are indicated
with bold symbols (i.e. ${\boldsymbol {\it A}}$) and three-vectors with the
standard arrow (i.e. ${\vec A}$). Finally, I will adopt geometrized units
in which $G=c=1$. All of the figures presented here are original and have
not been published elsewhere.

\subsection{Linearized Einstein Equations}

	The starting point in discussing gravitational waves cannot but
come from the Einstein field equations, expressing the close equivalence
between matter-energy and curvature
\begin{equation}
\label{efe1}
G_{\mu \nu} \equiv R_{\m \n} - \frac{1}{2} g_{\m \n} R = 
	8 \pi T_{\mu \nu} \ . 
\end{equation}
In the 10 linearly independent equations (\ref{efe1}), $R_{\m \n}$ and
$R$ are the Ricci tensor and scalar, respectively, $g_{\m \n}$ and $G_{\m
\n}$ are the metric and Einstein tensors, respectively, while $T_{\m \n}$
is the stress-energy tensor of the matter in the spacetime considered.

	Looking at the Einstein equations (\ref{efe1}) as a set of
second-order partial differential equations it is not easy to predict
that there exist solutions behaving as waves. Indeed, and as it will
become more apparent in this Section, the concept of gravitational waves
as solutions of Einstein equations is valid only under some rather
idealized assumptions such as: a vacuum and asymptotically flat spacetime
and a linearized regime for the gravitational fields. If these
assumptions are removed, the definition of gravitational waves becomes
much more difficult. In these cases, in fact, the full nonlinearity of
the Einstein equations complicates the treatment considerably and
solutions can be found only numerically. It should be noted, however,
that in this respect gravitational waves are not peculiar. Any wave-like
phenomenon, in fact, can be described in terms of exact ``wave
equations'' only under very simplified assumptions such as those requiring
an uniform ``background'' for the fields propagating as waves.

	These considerations suggest that the search for wave-like
solutions to Einstein equations should be made in a spacetime with very
modest curvature and with a metric line element which is that of flat
spacetime but for small deviations on nonzero curvature, i.e.
\begin{equation}
\label{metric}
g_{\mu \nu} = \eta_{\mu \nu} + h_{\mu \nu} + {\cal O}([h_{\m \n}]^2)\ ,
\end{equation}
where
\begin{equation}
\eta_{\m \n} = {\rm diag}(-1,1,1,1) \ ,
\end{equation}
and the linearized regime is guaranteed by the fact that 
\begin{equation}
\label{lin}
|h_{\mu \nu}| \ll 1 \ .
\end{equation}

	Fortunately, the conditions expressed by equations (\ref{metric})
and (\ref{lin}) are, at least in our Solar system, rather easy to
reproduce and, in fact, the deviation away from flat spacetime that could
be measured, for instance, on the surface of the Sun are
\begin{equation}
|h_{\m \n}| \sim |h_{00}| \simeq \frac{M_{\odot}}{R_{\odot}} 
	\sim 10^{-6} \ .
\end{equation}

	Before writing the linearized version of the Einstein equations
(\ref{efe1}) it is necessary to derive the linearized expression for the
Christoffel symbols. In a coordinate basis (as the one will will assume
hereafter), the general expression for the affine connection is
\begin{equation}
\Gamma^{\m}_{\ \al \bt} = \frac{1}{2}
	g^{\m \n}(g_{\n \al, \bt} + g_{\bt \n, \al} - g_{\al \bt, \n}) \ ,
\end{equation}
where the partial derivatives are readily calculated as 
\begin{equation}
g_{\n \al, \bt}  = \eta_{\n \al, \bt} + h_{\n \al, \bt}  
	= h_{\n \al, \bt} \ , 
\end{equation}
so that the linearized Christoffel symbols become
\begin{eqnarray}
\label{lin_gammas}
\Gamma^{\m}_{\ \al \bt} &=& \frac{1}{2}
	\eta^{\m \n}(h_{\n \al, \bt} + h_{\bt \n, \al} - h_{\al \bt, \n})
\nonumber \\
	&=& \frac{1}{2}
	(h^{\ \ \mu}_{\alpha \ \ , \bt} + 
	 h^{\ \ \mu}_{\beta  \ \ , \al} - 
	 h^{\ \ \ ,\mu}_{\alpha \beta}) \ .
\end{eqnarray}
Note that the operation of lowering and raising the indices in expression
(\ref{lin_gammas}) is not made through the metric tensors $g_{\m \n}$ and
$g^{\m \n}$ but, rather, through the spacetime metric tensors $\eta_{\m
\n}$ and $\eta^{\m \n}$. This is just the consequence of linearized
approximation and, despite this, the spacetime is really curved!

	Once the linearized Christoffel symbols have been computed, it is
possible to derive the linearized expression for the Ricci tensor which
takes the form
\begin{eqnarray}
\label{lin_rt}
R_{\m \n} &=& \Gamma^{\al}_{\ \m \n,\al} - \Gamma^{\al}_{\ \m \al,\n}
\nonumber \\
	&=& \frac{1}{2}
	(h^{\ \;\al}_{\m \ \;,\n \al} + h^{\ \;\al}_{\n \ \;,\,\m \al} -
	 h^{\ \ \ \ \;\al}_{\m \n,\, \al} - h_{,\,\m\n}) \ ,
\end{eqnarray}
where
\begin{equation}
h \equiv h^{\al}_{\ \; \al}  = \eta^{\m \al} h_{\m \al} \ ,
\end{equation}
is the trace of the metric perturbations. The resulting Ricci scalar is
then given by
\begin{equation}
\label{lin_rs}
R \equiv g^{\m \n} R_{\m \n}  \simeq \eta^{\m \n} R_{\m \n} \ .
\end{equation}

	Making now use of (\ref{lin_rt}) and (\ref{lin_rs}) it is
possible to rewrite the Einstein equations (\ref{efe1}) in a linearized
form as
\begin{equation}
\label{efe2}
h^{\ \ \ \al}_{\m \al,\ \n} + h^{\ \ \ \al}_{\n \al,\ \m} 
	- h^{\ \ \ \al}_{\m \n,\ \al} - h_{,\m\n} - 
	\eta_{\m \n}(h^{\ \ \ \al \bt}_{\al \bt ,} - h^{\ \ \al}_{,\al})
	= 16 \pi T_{\m \n} \ .
\end{equation}

	Although linearized, the Einstein equations (\ref{efe2}) do not
seem yet to suggest a wave-like behaviour. A good step in the direction
of unveiling this behaviour can be made if a more compact notation is
introduced and which makes use of {\it ``trace-free'' tensors} defined as
\begin{equation}
{\bar h}_{\m \n} \equiv h_{\m \n} - \frac{1}{2} \eta_{\m \n} h \ ,
\end{equation}
where the ``bar-operator'' can be applied to any symmetric tensor so
that, for instance, ${\bar R}_{\m \n} = G_{\m \n}$ and ${\bar {\bar
h}}_{\m \n} = h_{\m \n}$\footnote{Note that the ``bar'' operator can in
principle be applied also to the trace so that ${\bar h}=-h$}. Using this
notation, the linearized Einstein equations (\ref{efe2}) take the more
compact form
\begin{equation}
\label{efe3}
-{\bar h}^{\ \ \ \ \al}_{\m \n,\al} -\eta_{\m \n} 
	{\bar h}^{\ \ \ \al \bt}_{\al \bt,} + 
	{\bar h}^{\ \ \ \; \al}_{\n \al, \ \;\m}
	= 16 \pi T_{\m \n} \ .
\end{equation}

	It is now straightforward to recognize that the first term on the
right-hand-side of equation (\ref{efe3}) is simply the Dalambertian (or
wave) operator
\begin{equation}
\label{efe2.1}
{\bar h}^{\ \ \ \ \al}_{\m \n,\al} = \Box {\bar h}_{\m \n} = 
	-(- \partial^2_t + \partial^2_x + \partial^2_y + \partial^2_z )
	{\bar h}_{\m \n} \ ,
\end{equation}
where the last equality is valid for a Cartesian $(t,x,y,z)$ coordinate
system only. At this stage the gauge freedom inherent to General
Relativity can (and should) be exploited to recast equations
(\ref{efe2.1}) in a more convenient form. A good way of exploiting this
gauge freedom is by choosing the metric perturbations $h_{\m \n}$ so as
to eliminate the terms in (\ref{efe3}) that spoil the wave-like
structure. Most notably, the metric perturbations can be selected so that
\begin{equation}
\label{lg}
{\bar h}^{\m \al}_{\ \ \;,\al} = 0 \ .
\end{equation}
Making use of the gauge (\ref{lg}), which is also known as {\it
``Lorentz''} (or Hilbert) gauge, the linearized field equations take the
form
\begin{equation}
\label{efe4}
\Box{\bar h}_{\m \n} = - 16 \pi T_{\m \n} \ .
\end{equation}

\noindent Despite they are treated in a linearized regime and with a
proper choice of variables and gauges, Einstein equations (\ref{efe4}) do
not {\it yet} represent wave-like equations if matter is present (i.e. if
$T_{\m \n} \ne 0$). A further and final step needs therefore to be taken
and this amounts to consider a spacetime devoid of matter, in which the
Einstein equations can finally be written as
\begin{equation}
\label{efe5}
\Box{\bar h}_{\m \n} = 0  \ ,
\end{equation}
indicating that, in the Lorentz gauge, the ``gravitational field''
propagates in spacetime as a wave perturbing flat spacetime. 

	Having recast the Einstein field equations in a wave-like form
has brought us just half-way towards analysing the properties of these
objects. More will be needed in order to discuss the nature and features
of gravitational waves and this is what is presented in the following
Section.
	
\subsection{A Wave Solution to Einstein Equations}

	The simplest solution to the linearized Einstein equations
(\ref{efe5}) is that of a plane wave of the type
\begin{equation}
\label{sol_efe5}
{\bar h}_{\m \n} = {\Re \left \{ A_{\m \n} 
	\exp (i \kappa_{\al}x^{\al})\right \}} \ ,
\end{equation}
where ${\boldsymbol {\it A}}$ is the {\it ``amplitude tensor''} and
${\boldsymbol \kappa}$ is a null four-vector,
i.e. $\kappa^{\al}\kappa_{\al}=0$. In such a solution, the plane wave
(\ref{sol_efe5}) travels in the spatial direction ${\vec k} =
(\kappa_x,\kappa_y,\kappa_z)/\kappa^0$ with frequency $\omega \equiv
\kappa^0 = (\kappa^j\kappa_j)^{1/2}$.

	Note that the amplitude tensor ${\boldsymbol {\it A}}$ in the
wave solution (\ref{sol_efe5}) has in principle $16-6=10$ independent
components. On the other hand, a number of considerations indicate that
there are only two dynamical degrees of freedom in General
Relativity. This ``excess'' of independent components can be explained
simply. Firstly, ${\boldsymbol {\it A}}$ and ${\boldsymbol {\it \kappa}}$
cannot be arbitrary if they have to describe a plane wave; as a result,
an orthogonality condition between the two quantities will constrain four
of the ten components of ${\boldsymbol {\it A}}$ (see condition {\it (a)}
below). Secondly, while a global Lorentz gauge has been chosen
[cf. equation (\ref{lg})], this does {\it not} fix completely the
coordinate system of a linearized theory. A residual ambiguity, in fact,
is preserved through arbitrary {\it "gauge changes''}, i.e. through
infinitesimal coordinate transformations that are not entirely
constrained, even if a global gauge has been selected. To better
appreciate this, consider an infinitesimal coordinate transformation in
terms of a small but otherwise arbitrary displacement four-vector
${\boldsymbol {\it \xi}}$
\begin{equation}
\label{coord_trans}
x^{\al'} = x^{\al} + \xi^{\al} \ . 
\end{equation}
Applying this transformation to the linearized metric (\ref{metric})
generates a ``new'' metric tensor that at the lowest order is
\begin{equation}
\label{metric'}
g^{^{\rm NEW}}_{\mu'\nu'} = \eta_{\mu \nu} + h^{^{\rm OLD}}_{\mu \nu} 
	- \xi_{\m , \n} - \xi_{\n , \m}	\ ,
\end{equation}
so that the ``new'' and ``old'' perturbations are related by the
following expression
\begin{equation}
\label{hmn'}
h^{^{\rm NEW}}_{\mu'\nu'} = h^{^{\rm OLD}}_{\mu \nu} 
	- \xi_{\m , \n} - \xi_{\n , \m} \ .
\end{equation}
or, alternatively, by 
\begin{equation}
\label{hbmn'}
{\bar h}^{^{\rm NEW}}_{\mu'\nu'} = {\bar h}^{^{\rm OLD}}_{\mu \nu} -
	\xi_{\m , \n} - \xi_{\n , \m} +
	\eta_{\mu \nu} \xi^{\alpha}_{\ \ , \alpha}\ .
\end{equation}

	Requiring now that the new coordinates satisfy the condition
(\ref{lg}) of the Lorentz gauge, forces the displacement vector to be
solution of the sourceless wave equation
\begin{equation}
\label{we_xi}
\xi^{\alpha,\beta}_{\ \ \ \beta} = 0 \ . 
\end{equation}
As a result, the plane-wave vector with components
\begin{equation}
\label{xi_alpha}
\xi^{\alpha} \equiv -i C^{\alpha} {\rm exp}(i \kappa_{\beta}x ^{\beta}) 
	\ , 
\end{equation}
generates, through the four arbitrary constants $C^{\alpha}$, a gauge
transformation that changes arbitrarily four more components of
${\boldsymbol {\it A}}$. Effectively, therefore, $A_{\mu \nu}$ has only
$10-4-4=2$ linearly independent components, corresponding to the number
of degrees of freedom in General Relativity~\cite{mtw74}.

	Note that all this is very similar to what happens in classical
electrodynamics, where the Maxwell equations are invariant under
transformations of the vector potentials of the type $A_{\mu} \rightarrow
A_{\mu'} = A_{\mu} + \Psi_{,\m}$, so that the corresponding
electromagnetic tensor $F^{^{\rm NEW}}_{\m' \n'} = A_{\m',\n'} -
A_{\n',\m'} = F^{^{\rm OLD}}_{\m' \n'}$. Similarly, in a linearized
theory of General Relativity, the gauge transformation (\ref{hmn'}) will
preserve the components of the Riemann tensor, i.e. $R^{^{\rm
NEW}}_{\alpha \beta \m \nu} = R^{^{\rm OLD}}_{\alpha \beta \m \nu}$.

	In practice, it is often convenient to constrain the components
of the amplitude tensor through the following conditions:

\begin{description}

\item {\it (a)}: {\bf Orthogonality Condition}: {\it four} components of
the amplitude tensor can be specified if ${\boldsymbol {\it A}}$ and
${\boldsymbol \kappa}$ are chosen to be orthogonal
\begin{equation}
\label{oc}
A_{\m \n} \kappa^{\n} = 0 \ .
\end{equation}

\item {\it (b)}: {\bf Global Lorentz Frame}: just like in Special
Relativity, a global Lorentz frame relative to an observer with
four-velocity ${\boldsymbol {\it u}}$ can be defined. In this case, {\it
three}\footnote{Note that the conditions (\ref{oc}) fix three and not
four components because one further constraint needs to be satisfied,
i.e. $\kappa^{\m} A_{\m \n} u^{\n} = 0$.} components of the amplitude
tensor can be specified after selecting a four-velocity ${\boldsymbol
{\it u}}$ orthogonal to ${\boldsymbol {\it A}}$
\begin{equation}
A_{\m \n} u^{\n} = 0 \ .
\end{equation}

\item {\it (c)}: {\bf Infinitesimal Gauge Transformation}: {\it one}
final independent component in the amplitude tensor can be eliminated
after selecting the infinitesimal displacement vector $\xi^{\m} = iC^{\m}
\exp(i\kappa_{\al}x^{\al})$ so that

\begin{equation}
\label{igt}
A^{\m}_{\ \m} = 0 \ . 
\end{equation}

\end{description}

	Consider now the constraint conditions (\ref{oc})--(\ref{igt})
listed above as implemented in a reference frame which is globally at
rest, i.e. $u^{\al} = (1,0,0,0)$. In this frame, the components of the
wave vector $\kappa^{\m}$ do not appear directly, and the above
conditions for the amplitude tensor can be written as

\begin{description}

\item {\it (a):}
\begin{equation}
\label{(b)}
A_{\m \n} \kappa^{\n} = 0 \qquad \Longleftrightarrow 
	\qquad h_{ij,j } = 0 \ ,
\end{equation}
i.e. the spatial components of $h_{\m \n}$ are {\sl divergence-free}.

\item {\it (b):}
\begin{equation}
\label{(a)}
A_{\m \n} u^{\n} = 0 \qquad \Longleftrightarrow 
	\qquad h_{\m 0 } = 0 \ ,
\end{equation}
i.e. only the spatial components of $h_{\m \n}$ are {\sl nonzero}.

\item {\it (c):}
\begin{equation}
\label{(c)}
A^{\m}_{\ \m} = 0 \qquad \Longleftrightarrow 
	\qquad h=h^{j}_{\ j} = 0 \ ,
\end{equation}
i.e. the spatial components of $h_{\m \n}$ are {\sl trace-free}. Because
of this, and only in this gauge, ${\bar h}_{\m \n} = h_{\m \n}$

\end{description}

\noindent Conditions {\it (a), (b)} and {\it (c)} define the so called
{\it ``Transverse and Traceless'' (TT)} gauge and represent the standard
gauge for the analysis of gravitational waves.

	An obvious question that might emerge at this point is about the
generality of the $TT$ gauge. A simple answer to this question can be
provided by reminding that any linear gravitational wave can, just like
any electromagnetic wave, be decomposed in the linear superposition of
planar waves. Because all of the conditions (\ref{(a)})--(\ref{(c)}) are
linear in $h_{\m \n}$, any of the composing planar waves can be chosen to
satisfy (\ref{(a)})--(\ref{(c)}), which, as a result, are satisfied also
by the original gravitational wave. Indeed, all of what just stated is
contained in a theorem establishing that: once a global Lorentz frame has
been chosen in which $u^{\al} = \delta^{\al}_{\ 0}$, it is then always
possible to find a gauge in which the conditions (\ref{(a)})--(\ref{(c)})
are satisfied.

\subsection{Making Sense of the {\it TT} Gauge}	

	As introduced so far, the $TT$ gauge might appear rather abstract
and not particularly interesting. Quite the opposite, the $TT$ gauge
introduces a number of important advantages and simplifications in the
study of gravitational waves. The most important of these is that, in
this gauge, the only nonzero components of the Riemann tensor are
\begin{equation}
R_{j0k0}=R_{0j0k} = -R_{j00k} = -R_{0jk0}\ .
\end{equation}
Since, however, 
\begin{equation}
\label{rj0k0}
R_{j0k0}=-\frac{1}{2} \htt_{jk,00} \ ,
\end{equation}
the use of the $TT$ gauge indicates that a travelling gravitational wave
with periodic time behaviour $\htt_{jk} \propto \exp (i \omega t)$ can be
associated to a local oscillation of the spacetime, i.e. 
\begin{equation}
\htt_{jk,00}
 \sim -\omega^2 \exp (i \omega t) \sim R_{j0k0}\ , 
	\qquad {\rm and} \qquad \ 
	R_{j0k0} = \frac{1}{2} \omega^{2} \htt_{jk} \ .
\end{equation}

\begin{figure}[htb]
\begin{center}
\hspace{0.125truecm}
\leavevmode
\hbox{\psfig{figure=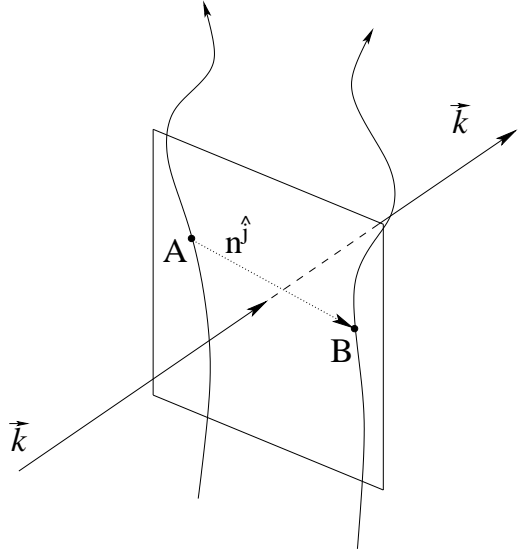,width=12.0truecm,angle=-90} }
\caption{\footnotesize Schematic diagram for the changes in the
separation vector between two particles $A$ and $B$ moving along geodesic
trajectories produced by the interaction with a gravitational wave
propagating along the direction ${\vec k}$.
\label{fig0}}
\end{center}
\end{figure}

	To better appreciate the effects of the propagation of a
gravitational wave, it is useful to consider the separation between two
neighbouring particles $A$ and $B$ on a geodesic motion and how this
separation changes in the presence of an incident gravitational wave (see
Figure~\ref{fig0}). For this purpose, let us introduce a coordinate
system $x^{\h \al}$ in the neighbourhood of particle $A$ so that along
the worldline of the particle $A$ the line element will have the form
\begin{equation}
ds^2 = -d\tau^2 + \delta_{\h i \h j} dx^{\h i} dx^{\h j} + {\cal
	O}(|x^{\h j}|^2) dx^{\h \al} dx^{\h \bt} \ .
\end{equation}
The arrival of a gravitational wave will perturb the geodesic motion of
the two particles and produce a nonzero contribution to the
geodesic-deviation equation. I remind that the changes in the separation
four-vector ${\boldsymbol {\it V}}$ between two geodesic trajectories
with tangent four-vector ${\boldsymbol {\it u}}$ are expressed through
the geodesic-deviation equation
\begin{equation}
u^{\gamma} u^{\bt} V^{\al}_{\ \ ;\bt \gamma} = 
	-R^{\al}_{\ \bt \gamma \delta} u^{\bt} V^{\gamma} u^{\delta}  \ ,
\end{equation}
or, equivalently, as
\begin{equation}
\label{gde}
u^{\gamma} u^{\bt} \left(\frac{D^2 V^{\al}}{D \tau^2} \right)
	\equiv u^{\gamma} u^{\bt} \left(
	\frac{d^2 V^{\al}}{d \tau^2} + \Gamma^{\al}_{\ \bt \gamma} 
	\frac{d V^{\al}}{d \tau}  \frac{d V^{\bt}}{d \tau} \right) = 
	-R^{\al}_{\ \bt \gamma \delta} u^{\bt} V^{\gamma}
	u^{\delta}\ . 
\end{equation}

	Indicating now with $n^{\h j}_{_{\rm B}} \equiv x^{\h j}_{_{\rm
B}} - x^{\h j}_{_{\rm A}} = x^{\h j}_{_{\rm B}}$ the components of the
separation three-vector in the positions of the two particles, the
geodesic-deviation equation (\ref{gde}) can be written as
\begin{equation}
\label{gde1}
\frac{D^2 x^{\h j}_{_{\rm B}}}{D \tau^2} =
	-R^{\h j}_{\ 0 \h k 0} x^{\h k}_{_{\rm B}} \ . 
\end{equation}
A first simplification to these equations comes from the fact that around
the particle $A$ the affine connections vanish (i.e. $\Gamma^{\h
j}_{{\hat \al} {\hat \bt}}=0$) and the covariant derivative in
(\ref{gde1}) can be replaced by an ordinary total derivative.
Furthermore, because in the $TT$ gauge the coordinate system $x^{\hat
\al}$ moves together with the particle $A$, the proper and the coordinate
time coincide at first order in the metric perturbation [i.e. $\tau=t$ at
${\cal O}(h^{^{\rm TT}}_{\m \n})$]. As a result, equation (\ref{gde1})
effectively becomes
\begin{equation}
\label{gde2}
\frac{d^2 x^{\hat j}_{_{\rm B}}}{d t^2} =
	\frac{1}{2}\left(\frac{\partial^2 \htt_{{\hat j} {\hat k}}}
	{\partial t^2} \right) x^{\hat k}_{_{\rm B}} \ ,
\end{equation}
and has solution
\begin{equation}
\label{gde_sol}
x^{\hat j}_{_{\rm B}}(t) = x^{\hat k}_{_{\rm B}}(0) \left[
	\delta_{{\hat j} {\hat k}} + \frac{1}{2} 
	\htt_{{\hat j} {\hat k}}(t)\right] \ .
\end{equation}
Equation (\ref{gde_sol}) has a straightforward interpretation and
indicates that, in the reference frame comoving with $A$, the particle
$B$ is seen oscillating with an amplitude proportional to $\htt_{{\hat j}
{\hat k}}$.

	Note that because these are transverse waves, they will produce a
local deformation of the spacetime only in the plane orthogonal to their
direction of propagation. As a result, if the two particles lay along the
direction of propagation (i.e. if ${\vec n} \parallel {\vec k}$), then
$\htt_{{\hat j} {\hat k}} x^{\hat j}_{_{\rm B}}(0) \propto \htt_{{\hat j}
{\hat k}} \kappa^{\hat j}_{_{\rm B}}(0) = 0$ and no oscillation will be
recorded by $A$ [cf. equation (\ref{(b)})]

\begin{figure}[htb]
\begin{center}
\leavevmode
\hbox{\psfig{figure=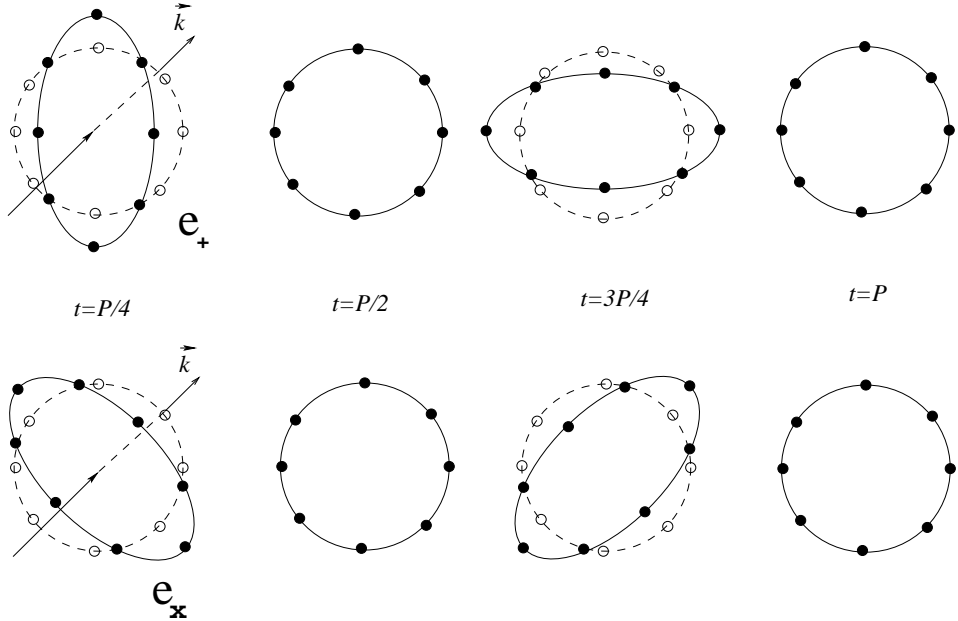,width=14.0truecm,angle=-90} }
\caption{\footnotesize Schematic deformations produced on a ring of
freely-falling particles by gravitational waves that are linear polarized
in the ``$+$'' (``plus'') and ``$\times$'' (``cross'') modes. The
continuous lines and the dark filled dots show the positions of the
particles at different times, while the dashed lines and the open dots
show the unperturbed positions.\label{fig1}}
\end{center}
\end{figure}
	Let us now consider a concrete example and in particular a planar
gravitational wave propagating in the positive $z$-direction. In this
case
\begin{eqnarray}
\htt_{xx} &=& - \htt_{yy} = {\Re  \left\{ A_{+} 
	\exp[-i\omega(t-z)]\right\} } \ ,
	\\ \nonumber \\
\htt_{xy} &=& \htt_{yx} = {\Re \left\{ A_{\times}
	\exp[-i\omega(t-z)]\right\}} \	, 
\end{eqnarray}
\noindent where $A_{+}$ and $A_{\times}$ represent the two independent
modes of polarization. 

\noindent As in classical electromagnetism, in fact, it is possible to
decompose a gravitational wave in two {\it linearly} polarized plane
waves or in two {\it circularly} polarized ones. In the first case, and
for a gravitational wave propagating in the $z$-direction, the
polarization {\sl tensors} $+$ (``plus'') and $\times$ (``cross'') are
defined as
\begin{eqnarray}
{\mathbf e}_{+} &\equiv& {\vec e}_{x} \otimes 
	{\vec e}_{x} - {\vec e}_{y} 
	\otimes {\vec e}_{y} \ ,
\\ \nonumber \\
{\mathbf e}_{\times} &\equiv& {\vec e}_{x} \otimes 
	{\vec e}_{x} + {\vec e}_{y} 
	\otimes {\vec e}_{y} \ .
\end{eqnarray}

	The deformations that are associated with these two modes of
linear polarization are shown in Figure~\ref{fig1} where the positions of
a ring of freely-falling particles are schematically represented at
different fractions of an oscillation period. Note that the two linear
polarization modes are simply rotated of $\pi/4$.

\begin{figure}[htb]
\begin{center}
\leavevmode
\hbox{\psfig{figure=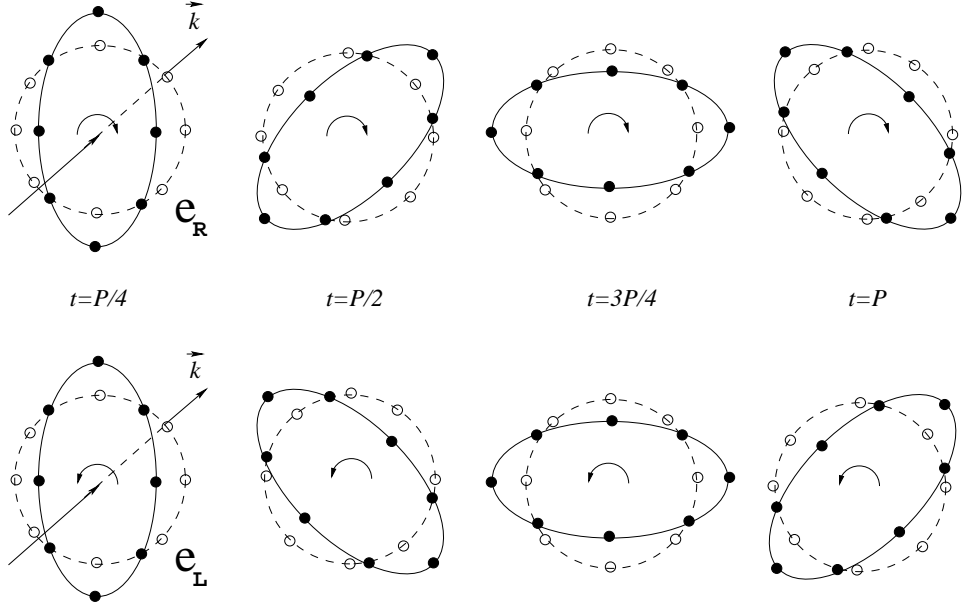,width=14.0truecm,angle=-90} }
\caption{\footnotesize Schematic deformations produced on a ring of
freely-falling particles by gravitational waves that are circularly
polarized in the $R$ (clockwise) and $L$ (counter-clockwise) modes. The
continuous lines and the dark filled dots show the positions of the
particles at different times, while the dashed lines and the open dots
show the unperturbed positions. \label{fig2}}
\end{center}
\end{figure}

	In a similar way, it is possible to define two {\sl tensors}
describing the two states of circular polarization and indicate with
${\bf e}_{_{\rm R}}$ the circular polarization that rotates clockwise
(see Figure~\ref{fig2})
\begin{equation}
{\bf e}_{_{\rm R}} \equiv \frac{{\bf e}_{+} + i 
	{\bf e}_{\times}}{\sqrt 2} \ ,
\end{equation}
and with ${\bf e}_{_{\rm L}}$ the circular polarization that rotates
counter-clockwise (see Figure~\ref{fig2})
\begin{equation}
{\bf e}_{_{\rm L}} \equiv \frac{{\bf e}_{+} - i
	{\bf e}_{\times}}{\sqrt 2} \ .
\end{equation}
The deformations that are associated to these two modes of circular
polarization are shown in Figure~\ref{fig2}

\subsection{Generation of Gravitational Waves}

In what follows I briefly discuss the amounts of energy carried by
gravitational waves and provide simple expressions to estimate the
gravitational radiation luminosity of potential sources. Despite the
estimates made here come from analogies with electromagnetism and are
based on a Newtonian description of gravity, they provide a reasonable
approximation to more accurate expressions from which they differ for
factors of ${\cal O}({\rm few})$. Note also that if obtaining such a
level of accuracy requires only a small effort, reaching the accuracy
necessary for a realistic detection of gravitational waves is far more
difficult and often imposes the use of numerical relativity calculations
on modern supercomputers.

	In classical electrodynamics, the energy emitted per unit time by
an oscillating electric dipole $d$ is easily estimated to be
\begin{equation}
\label{ed}
L_{\rm electric\ dip.} \equiv \frac{(\rm energy\ emitted)}{(\rm unit\ time)}
	=\frac{2}{3}q^2 a^2 =\frac{2}{3}({\ddot d})^2 \ ,
\end{equation}
where 
\begin{equation}
d\equiv q x \ , \qquad {\rm and} \qquad {\ddot d} \equiv q {\ddot x} \ , 
\end{equation}
with $q$ being the electrical charge and the number of ``dots'' counting
the order of the total time derivative. Equally simple is to calculate
the corresponding luminosity in gravitational waves produced by an
oscillating mass-dipole. In the case of a system of $N$ point-like
particles of mass $m_A$ ($A=1,2,\ldots,N$), in fact, the total
mass-dipole and its first time derivative are 
\begin{equation}
{\vec d} \equiv \sum^N_{A=1} m_A {\vec x}_A \ ,
\end{equation}
and
\begin{equation}
{\dot {\vec d}} \equiv \sum^N_{A=1} m_A {\dot {\vec x}}_A = 
	{\vec p}\ ,
\end{equation}
respectively. However, the requirement that the system conserves its
total linear momentum
\begin{equation}
{\ddot {\vec d}} \equiv {\dot {\vec p}}=0\ , 
\end{equation}
forces to conclude that $L_{\rm mass\ dipole}=0$, i.e. that there is no
mass-dipole radiation in General Relativity (This is equivalent to the
impossibility of having electromagnetic radiation from an oscillating
electric monopole.). Next, consider the electromagnetic energy emission
produced by an oscillating electric quadrupole. In classical
electrodynamics, this energy loss is given by
\begin{equation}
\label{eq}
L_{\rm electric\ quad.} \equiv 
	\frac{1}{20}({\dddot Q})^2 =\frac{1}{20}
	({\dddot Q_{jk}}{\dddot Q_{jk}})^2 \ ,
\end{equation}
where	
\begin{equation}
Q_{jk} \equiv \sum^N_{A=1} q_A \left[ (x_A)_j (x_A)_k - 
	\frac{1}{3}\delta_{jk} (x_A)_i (x_A)^i\right]\ ,
\end{equation}
is the electric quadrupole for a distribution of $N$ charges $(q_1, q_2,
\ldots, q_N)$.

\noindent In close analogy with expression (\ref{eq}), the energy loss
per unit time due to an oscillating mass quadrupole is calculated to be
\begin{equation}
\label{mq}
L_{\rm mass\ quadrupole} \equiv 
	\frac{1}{5}\frac{G}{c^5}\langle\; 
	{\dddot {\!\boldsymbol I} \; }\rangle^2 =
	\frac{1}{5}\frac{G}{c^5}\langle {\dddot \Itf_{jk}}
	{\dddot \Itf_{jk}} \rangle^2 \ ,
\end{equation}
where $\Itf_{jk}$ is the trace-less mass quadrupole (or ``reduced'' mass
quadrupole), defined as
\begin{eqnarray}
\label{itf}
\Itf_{jk} &\equiv& \sum^N_{A=1} m_A \left[ (x_A)_j (x_A)_k - 
	\frac{1}{3}\delta_{jk} (x_A)_i (x_A)^i\right] 
\nonumber \\
	&=& 
	\int \rho \left(x_j x_k - \frac{1}{3} 
	\delta_{jk} x_i x^i\right) dV\ ,
\end{eqnarray}
and the brackets $\langle ~ \rangle$ indicate a time average [Clearly,
the second expression in (\ref{itf}) refers to a continuous distribution
of particles with rest-mass density $\rho$.].

\noindent A crude estimate of the third derivative of the mass quadrupole
of the system is given by
\begin{equation}
\label{rough}
{\dddot \Itf_{jk}} \sim \frac{(\rm mass\ of\ the\ system\ in\ motion)
	\times (size\ of\ the\ system)^2}{(\rm time\ scale)^3} 
	\sim \frac{M R^2}{\tau^3} \sim 
	\frac{M v^2}{\tau} \ ,
\end{equation}
so that
\begin{equation}
\label{crude}
L_{\rm mass\ quadrupole} \sim
	\frac{G}{c^5} \frac{M v^2}{\tau} \ .
\end{equation}

	Although extremely simplified, expressions (\ref{rough}) and
(\ref{crude}) contain the two most important pieces of information about
the generation of gravitational waves. The first one is that the energy
emission in gravitational waves is severely suppressed by the coefficient
$G/c^5 \sim 10^{-59}$ or, stated differently, that the conversion of any
type of energy into gravitational waves is, in general, not
efficient. The second one is about the time variation of the mass
quadrupole, which can become considerable only for very {\it large
masses} moving at {\it relativistic speeds}. Clearly, these conditions
cannot be reached by sources in terrestrial laboratories, but can be
easily met by astrophysical compact objects, which therefore become the
most promising sources of gravitational radiation.

\subsubsection{Astrophysical Sources of Gravitational Waves}

	The research area that is involved with modelling the
astrophysical sources of gravitational waves is quite vast and is
increasing steadily as gravitational detectors become more and more
sensitive, and as detectors of new generation are becoming operative.
Having a detailed discussion of the multiple aspects of this line of
research, which encompasses numerical and perturbative techniques, is
beyond the scope of these lectures. However, the interested reader will
find more detailed discussions in the following review
articles~\cite{ct01, h02}.

\newpage
\section{Gravitational Waves From Perturbed Black Holes}

	This Section is dedicated to the analysis of the perturbations
that characterise a black hole and more precisely a nonrotating (or
Schwarzschild) black hole. Hereafter I will assume that the reader is
familiar with the basic properties of such a black hole as a static
solution to Einstein equations in a spherically symmetric and vacuum
spacetime. More precisely, I will consider as known the fundamental
concepts about physical and coordinate singularities, about the existence
of an event horizon as well as Birkhoff's theorem on the uniqueness of
Schwarzschild solution. All of these concepts are clearly discussed
in~\cite{mtw74,s84,di90}.

	Before discussing in detail black hole perturbations, one might
wonder why black hole perturbations are interesting at all. Indeed, there
are a number of good reasons why it is interesting and important to
consider black hole perturbations. Firstly, the presence of perturbations
can break the static properties of a black hole spacetime and be
therefore at the origin of gravitational wave {\it emission}. Secondly,
the gravitational waves emitted by a black hole carry information about
its {\it properties} such as mass, spin and charge. Finally, by
investigating the response of black holes to perturbations it is possible
to deduce important conclusions on the {\it stability} of these objects
(being a solution of Einstein equations, in fact, is just a sufficient
condition for stability).

	Because this is such an important area of research, some of the
main results date back to the first studies made by Regge and Wheeler
\cite{rw57} and the subsequent developments that took place in the 70's
\cite{z70, v70, m74}.

\subsection{Linear Perturbations of Black Holes}
\label{pobhs}

	The starting point in the analysis of black hole perturbations
is, of course, the unperturbed solution represented by a Schwarzschild
black hole with line element
\begin{equation}
\label{dsz}
ds^{2}=\gzdmn dx^{\mu}dx^{\nu} \equiv -\left( 1-\frac{2M}{r}\right)
	dt^{2}+\left( 1-\frac{2M}{r}\right)^{-1}dr^{2}+r^{2}d\Omega^{2}
	\ ,
\end{equation}
and where $\gzdmn$ represents the metric tensor of the static background
spacetime. Because the latter is assumed to be vacuum, the Einstein
equations assume a more compact form and can be written as
\begin{equation}
\label{efe_v}
\Rzdmn = 0 \ ,
\end{equation}
where $\Rzdmn$ is the Ricci tensor built with the background metric
$\gzdmn$. If small perturbations $h_{\m \n}$ are now introduced, the
resulting metric will be
\begin{equation}
g_{\mu \nu}=\gzdmn + h_{\mu \nu} \ ,
\end{equation}
where, again, the perturbations considered are much smaller than the
background, i.e.
\begin{equation}
|h_{\mu \nu}|/|\gzdmn| \ll 1 \ .
\end{equation}

	Just as for the static background, the behaviour of the perturbed
spacetime will be expressed by the Einstein equations that, using the
same notation as in (\ref{efe_v}), can be written as
\begin{equation}
\label{efe_vp}
R_{\mu \nu} =0 \ ,
\end{equation}
where $R_{\mu \nu} = R_{\m \n} (g_{\mu \nu})$. At first order in the
perturbations, the linearity can be exploited to break up the Einstein
equations (\ref{efe_vp}) as
\begin{equation}
\Rzdmn + \delta R_{\mu \nu} = 0 \ ,
\end{equation}
where $\delta R_{\mu \nu} \equiv R_{\m \n} (h_{\mu \nu})$. Using now
equations (\ref{efe_v}), the field equations reduce to
\begin{equation}
\label{efe_p}
\delta R_{\mu \nu} = 0 \ .
\end{equation}
	A different way of writing equations (\ref{efe_p}) is in terms of the
Christoffel symbols and more precisely as

\begin{equation}
\delta R_{\mu \nu}= - \delta \Gamma_{\mu \nu ;\beta}^{\beta} + 
	\delta \Gamma_{\mu \beta ;\nu}^{\beta} = 0\ ,
\end{equation}
where the perturbed Christoffel symbols are defined as
\begin{equation}
\delta \Gamma_{\mu \nu}^{\beta} = \frac{1}{2} \gzuab
	\left( h_{\mu \alpha ,\nu} + 
	h_{\nu \alpha ,\mu}-h_{\mu \nu ,\alpha}\right) \ .
\end{equation}

	Before proceeding further I should comment on some of the
properties of the metric perturbations $h_{\m \n}$. An important
constraint is posed by Birkhoff's theorem, which states that the
Schwarzschild solution is the only spherically symmetric, asymptotically
flat solution of Einstein equations in vacuum even if the spacetime is
not static~\cite{mtw74,s84,di90}. As a result, nonrotating black holes
can only be perturbed by nonradial perturbations and this forces to
consider perturbations with complete angular dependence, i.e.  $h_{\mu
\nu}= h_{\mu \nu}\left( t, r, \theta, \phi \right)$. Handling a generic
angular dependence can be complicated, but the mathematical treatment can
be simplified if the tensor perturbations $h_{\mu \nu}$ are written in a
separable form i.e. as the product of four parts each being a function of
one coordinate only.

\noindent In the case of a {\sl scalar function} depending on the spatial
coordinates only, it is well-known that this can be done after expanding
it in a series of {\it spherical harmonic functions}
\begin{equation}
\label{sh}
f\left( r,\theta, \phi \right) = 
	\sum_{l,m} a_{lm}(r) Y_{lm}(\theta, \phi) \ .
\end{equation}
In a similar way, in the case of a {\sl vector}, the separability is
achieved through an expansion in a series of {\it vector spherical
harmonics}~\cite{t80}
\begin{equation}
\label{vsh}
V^{\al}(r,\theta,\phi) = \sum_{l,m} a_{lm}(r)
	[Y^{\rm	B}_{lm}(\theta,\phi)]^{\al} +  \sum_{l,m}
	b_{lm}(r) [Y^{\rm	E}_{lm}(\theta,\phi)]^{\al}\ , 
\end{equation}
where $Y^{\rm B}_{lm}(\theta,\phi)$ and $Y^{\rm E}_{lm}(\theta,\phi)$ are
vector spherical harmonics of magnetic (B) and electric (E) type,
respectively. It will not therefore surprise that a series expansion of
the type (\ref{sh}) and (\ref{vsh}) can be used also for a rank-2
symmetric {\sl tensor} which can then be expanded in a series of {\it
tensor spherical harmonics}
\begin{equation}
\label{tsh}
T_{\m \n}(t,r,\theta,\phi) = 
	\sum_{l,m} a_{lm}(t,r)\left[A^{\rm ax}_{lm}
		(\theta,\phi)\right]_{\m \n} + 
	\sum_{l,m} b_{lm}(t,r)\left[B^{\rm pol}_{lm}
	(\theta,\phi)\right]_{\m \n} 
	\ , 
\end{equation}
where attention has been paid to the fact that, in general, a rank-2
symmetric tensor can be expanded in terms of tensor spherical harmonics
that behave differently under parity transformation, i.e. $(A^{\rm
ax}_{lm})_{\m \n}$ and $(B^{\rm pol}_{lm})_{\m \n}$.

	More specifically, if ${\cal P}$ is the {\it parity operator},
that is an operator producing a parity transformation on a rank-2
symmetric tensor $F_{\m \n}$
\begin{equation}
{\cal P}\left(\left[ F_{lm}\left( \theta, \phi \right) 
	\right]_{\mu \nu}\right) \ \
	\longrightarrow \ \ 
	\left[\widetilde{F}_{lm}\left(\pi - \theta, \pi + \phi \right) 
	\right]_{\mu \nu}\ ,
\end{equation}
the tensor spherical harmonics can then be classified according to their
behaviour ``under parity change''. In practice, are referred to as {\it
odd} or {\it axial} (or sometimes {\it toroidal}) those tensor harmonics
for which ${\cal P}\left(F_{\mu \nu}\right) = {\widetilde F}_{\m \n} =
\left( -1\right)^{l+1}F_{\mu \nu}$. Similarly, are referred to as {\it
even} or {\it polar} (or sometimes {\it spheroidal}) those tensor
harmonics for which ${\cal P}\left(F_{\mu \nu}\right) = {\widetilde
F}_{\m \n} = \left( -1\right)^{l}F_{\mu \nu}$.

	This classification of the tensor spherical harmonics is
reflected also on the metric perturbations that, as a result, are
classified as ``odd'' and ``even-parity'' respectively. However, before
discussing the specific forms that the perturbations assume in these
cases, it is convenient to introduce the so-called ``$3+1$''
decomposition, in which the spacetime is ``sliced'' into a family of
space-like spatial hypersurfaces parametrized by a $t = {\rm const.}$
coordinate. With this decomposition the line element takes the form
\begin{eqnarray}
ds^{2} &=& -\alpha^2 dt^{2}+\gamma_{ij}( dx^{i}+\beta^{i} dt)
	(dx^{j} + \beta^{j}dt) 
\nonumber \\ \nonumber \\
	&=& -(\alpha^2 - \beta^j\beta_j)dt^{2} + 2\beta_i dx^i dt + 
	\gamma_{ij} dx^{i}dx^{j} \ ,
\end{eqnarray}
where $\alpha$ is the {\it ``lapse''} function (expressing the rate at
which clocks on different hypersufaces tick) and $\bt^i$ are the components
of the {\it ``shift''} vector (relating coordinate changes on two
different hypersurfaces). With this decomposition, the metric
perturbations are then expressed in terms of a purely time part
($h_{00}$), of a purely spatial part ($h_{ij}$), and of a mixed
time-space part ($h_{0i}$)
\begin{equation}
\label{hmn_3po}
h_{\mu \nu}=\left( 
\begin{tabular}{cc}
$h_{00}$ & $h_{i0}$ \\ 
$h_{0i}$ & $h_{ij}$
\end{tabular}
\right) \ .
\end{equation}
In what follows I discuss the basic expressions of $h_{00}, h_{0j}$ and
$h_{ij}$ for the two classes of perturbations as well as the equations
they satisfy.

\subsection{Odd-parity Perturbations: the Regge-Wheeler Equation}

	I first consider the parts of the metric perturbations that are
of ``odd-parity'' [i.e. the first terms in the decomposition given in
equation (\ref{tsh})]. In this case, it is customary to introduce the
unknown functions $h_0(t,r), h_1(t,r)$ and $h_2(t,r)$ so that the
components of (\ref{hmn_3po}) can be written as
\begin{eqnarray}
\label{hmn_odd1}
h_{00} &=&0 \ ,\\ \nonumber \\
\label{hmn_odd2}
h_{0i} &=&h_{0}\left( t,r \right) \left[ 0,-\frac{1}{\sin \theta }%
	\sum_{l,m}\partial_{\phi }Y_{lm},\sin \theta 
	\sum_{l,m}\partial_{\theta }Y_{lm}\right] \ ,\\ \nonumber \\
\label{hmn_odd3}
h_{ij} &=&h_{1}\left( t,r \right) \left( \hat{e}_{1}\right)_{ij}
	+h_{2}\left( t,r \right) \left( \hat{e}_{2}\right)_{ij}\ ,
\end{eqnarray}
where $(\hat{e}_{1,2})_{ij}=\sum_{l,m} \left[ (\hat{e}_{1,2})_{ij}
\right]_{lm}$ and I will hereafter omit the $l,m$ indices and the sum over
them to maintain the expressions compact. The tensor spherical harmonics
$(\hat{e}_{1,2})_{ij}$ in (\ref{hmn_odd1})--(\ref{hmn_odd3}) have rather
lengthy but otherwise straightforward expressions which are given by
\begin{equation}
\left( \hat{e}_{1}\right)_{ij}=\left(
\begin{tabular}{ccc}
$0$ & $\displaystyle -\frac{1}{\sin \theta }\partial_{\phi }Y_{lm}$ & 
	$\sin \theta \partial_{\theta }Y_{lm}$ \\ \\
$\displaystyle -\frac{1}{\sin \theta }\partial_{\phi }Y_{lm}$ & 
	$0$ & $0$ \\ \\

$\sin \theta \partial_{\theta }Y_{lm}$ & $0$ & $0$%
\end{tabular}
\right) \ ,
\end{equation}
and
\begin{equation}
\label{e2ij}
\left( \hat{e}_{2}\right)_{ij}=\left( 
\begin{tabular}{ccc}
$0$ & $0$ & $0$ \\ \\
$0$ & $\displaystyle \frac{1}{\sin \theta }
	\left( \partial_{\theta \phi }^{2}-\cot \theta
	\partial_{\phi }\right) Y_{lm}$ 
	& $\displaystyle\frac{1}{2}\left[ 
	\displaystyle\frac{1}{\sin^{2}\theta }
	\partial_{\phi }^{2}-\cos \theta \partial_{\theta }-
	\sin\theta \partial_{\theta }^{2}\right] Y_{lm}$ \\ \\
	$0$ & $\displaystyle\frac{1}{2}\left[ 
	\displaystyle\frac{1}{\sin^{2}\theta }\partial_{\phi}^{2} - 
	\cos \theta \partial_{\theta }-\sin \theta \partial_{\theta }^{2}%
	\right] Y_{lm}$ & 
	$-\left[ \sin \theta \partial_{\theta \phi }^{2}-
	\cos\theta \partial_{\phi }\right] Y_{lm}$%
\end{tabular}
\right) \ .
\end{equation}
	
	The Einstein equations with the metric perturbations
(\ref{hmn_odd1})--(\ref{e2ij}) can be simplified if suitable gauge
conditions are chosen. I remind, in fact, that because of the linearized
approach, any infinitesimal coordinate transformation will lead to new
metric perturbations that are determined after the specification of the
suitable conditions for the displacement four-vector $\xi^{\m}$
[cf. equations (\ref{coord_trans})--(\ref{hmn'})].  While these
conditions are totally arbitrary, it is convenient to choose those
producing a simplification of the equations and, in the case of odd
perturbations, the choice usually made is that
\begin{equation}
h_{2}\left( t,r \right) = 0 \ . 
\end{equation}
In this gauge, which is usually referred to as the {\it ``Regge-Wheeler''
gauge}~\cite{rw57}, the odd-parity metric perturbations assume the
simplified form
\begin{equation}
\label{hmn_rwg}
h_{\mu \nu}^{\rm ax}=\left( 
\begin{tabular}{cccc}
$0$ & $0$ & $0$ & $h_{0}$ \\ 
$0$ & $0$ & $0$ & $h_{1}$ \\ 
$0$ & $0$ & $0$ & $0$ \\ 
$h_{0}$ & $h_{1}$ & $0$ & $0$%
\end{tabular}
\right) \sin \theta \partial_{\theta }P_{l}\left( \cos \theta \right) 
	e^{im \phi}\ ,
\end{equation}
where $P_{l}\left( \cos \theta \right)$ is the Legendre polynomial of
order $l$. Besides being simpler, Einstein equations in this gauge are
independent of $m$ in the sense that the final result will not depend on
the specific value chosen for $m$, which can therefore set to be zero. As
a result, the Einstein equations for the perturbed metric (\ref{hmn_rwg})
lead to the following system of equations
\begin{eqnarray}
\label{rw1}
&&\frac{\partial^{2}Q}{\partial t^{2}}-\frac{\partial^{2}Q}{\partial
	r_*^{2}}+\left( 1-\frac{2M}{r}\right) \left[ 
	\frac{l\left( l+1\right)}{r^{2}}- \frac{6M}{r^{3}}\right] Q = 0 \ ,
\\ \nonumber \\
\label{rw2}
&&\frac{\partial h_{0}}{\partial t }=\frac{\partial }{\partial r_*}
	\left(r_* Q\right) \ ,
\end{eqnarray}
where
\begin{equation}
Q \equiv \frac{h_{1}}{r}\left( 1-\frac{2M}{r}\right) \ ,
\end{equation}
and 
\begin{equation}
r_*\equiv r+2M\ln \left( \frac{r}{2M}-1\right) \ ,
\end{equation}
is the {\it ``tortoise coordinate''}. Because $r_*\rightarrow r$ for
$r\rightarrow \infty$ and $r_*\rightarrow -\infty$ as $r\rightarrow
2M^+$, the tortoise coordinate is particularly suited to study the
propagation of perturbations near the black hole event horizon which, in
this coordinate system, is placed at $-\infty$ and therefore does not
suffer from coordinate singularities.

\begin{figure}[htb]
\begin{center}
\leavevmode
\hbox{\psfig{figure=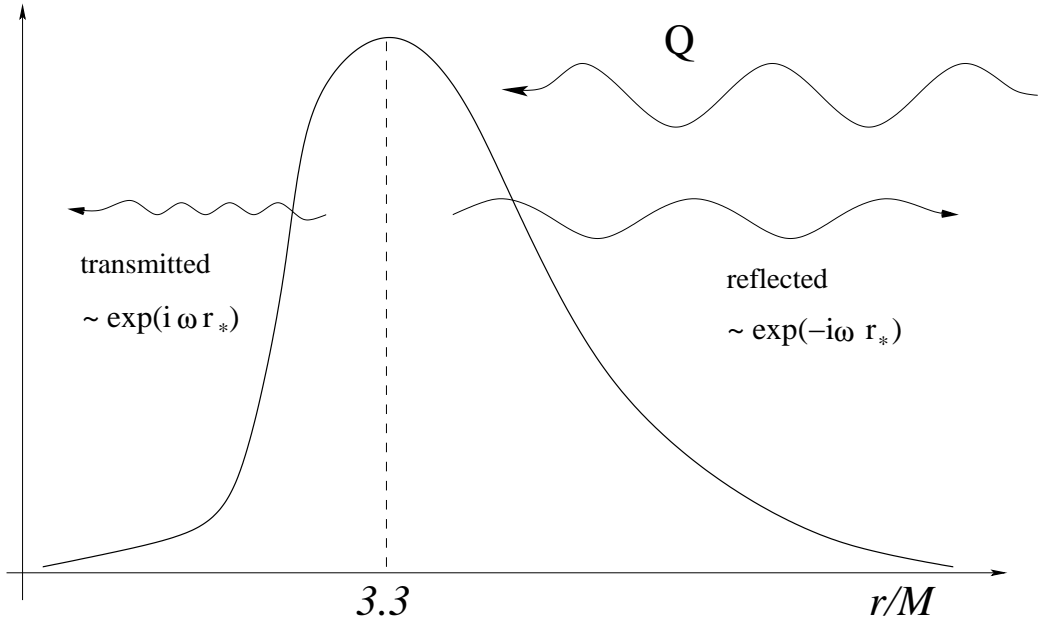,width=15.0truecm,angle=-90} }
\caption{\footnotesize Schematic representation of the Regge-Wheeler
potential $V$ [cf. eq. (\ref{rwp})] and of its effect on a incident
perturbation that is partially transmitted and partially reflected.
\label{fig3}}
\end{center}
\end{figure}

	A convenient way of looking at the ``Regge-Wheeler equation''
(\ref{rw1}) is that of considering it as a wave equation in a scattering
potential barrier $V(r)$, where
\begin{equation}
\label{rwp}
V(r)\equiv\left( 1-\frac{2M}{r}\right) \left[ 
	\frac{l\left( l+1\right) }{r^{2}} + p \right] \ ,
	\ \qquad {\rm with}
	\qquad p = -\frac{6M}{r^{3}} \ .
\end{equation}
The potential (\ref{rwp}) is also referred to as the ``Regge-Wheeler
potential'' and has a maximum just outside the event horizon, at $r \sim
3.3 M$ in Schwarzschild coordinates, as shown in the schematic
representation in Figure~\ref{fig3}. 

	In this view, the Regge-Wheeler equation shares all of the
well-known properties of a wave equation in a scattering potential and
the numerous results that have been found for this type of equation
(cf. Schr\"odinger equation) can also be applied to the propagation of
perturbations in the spacetime of Schwarzschild black hole
(see~\cite{c92} for a detailed discussion).
	
	As an example, a metric perturbations reaching the black hole
from spatial infinity can be regarded as a wave packet that will scatter
against the potential barrier $V$. As in quantum mechanics, not all of
the wave packet will be transmitted through the potential and some of it,
depending on the properties of the packet itself, will be reflected and
reach again spatial infinity (see also Figure~\ref{fig3}). This is
different from what happens, for instance, with a spherical massive shell
falling radially onto the black hole. These radical differences underline
the importance of a perturbative analysis of black hole spacetimes.

	Another interesting aspect of the Regge-Wheeler equation is that
it holds in a similar form also for scalar and vector perturbations, with
the only difference appearing in the effective potential (\ref{rwp}),
where $p=2M/r^3$ for scalar perturbations and $p=0$ for vector
ones~\cite{ks99}.

\subsection{Even-parity Perturbations: the Zerilli Equation}

	Next, I consider metric perturbation that are ``even-parity'' (or
polar). The mathematical approach is similar to the one followed for the
odd-parity perturbations and also in this case it is useful to introduce
a number of unknown functions $h_0(t,r), h_1(t,r), H_0(t,r), H_1(t,r),
H_2(t,r),\break K(t,r)$ and $G(t,r)$ so that the perturbed metric
functions can be written as
\begin{eqnarray}
h_{00} &=&-\frac{1}{2}\left( 1-\frac{2M}{r}\right)^{1/2}H_{0}
	\left(t,r\right) Y_{lm} \ , \\ \nonumber \\
h_{0i} &=&\left[ H_{1}Y_{lm},h_{0}\partial_{\theta }Y_{lm},h_{0}
	\partial_{\phi }Y_{lm}\right] \ , \\ \nonumber \\
h_{ij} &=&h_{1}\left( \hat{f}_{1}\right)_{ij}+\frac{H_{2}}{1-
	2M/r} \left( \hat{f}_{2}\right)_{ij}+r^{2} K
	\left( \hat{f}_{3}\right)_{ij}+
	r^{2}G\left( \hat{f}_{4}\right)_{ij} \ ,
\end{eqnarray}
where the tensor spherical harmonics $(\hat{f}_{1-4})_{ij}$ have the
forms
\begin{equation}
\left( \hat{f}_{1}\right)_{ij}=\left( 
\begin{tabular}{ccc}
$0$ & $\partial_{\theta }Y_{lm}$ & $\partial_{\phi }Y_{lm}$ \\ \\
$\partial_{\theta }Y_{lm}$ & $0$ & $0$ \\ \\
	$\partial_{\phi }Y_{lm}$ & $0$ & $0$%
\end{tabular}
\right)\ ,
\end{equation}
\begin{equation}
\left( \hat{f}_{2}\right)_{ij}=\left( 
\begin{tabular}{ccc}
$Y_{lm}$ & $0$ & $0$ \\ \\
$0$ & $0$ & $0$ \\ \\
$0$ & $0$ & $0$%
\end{tabular}
\right) \ ,
\end{equation}
\begin{equation}
\left( \hat{f}_{3}\right)_{ij}=\left( 
\begin{tabular}{ccc}
$0$ & $0$ & $0$ \\ \\
$0$ & $Y_{lm}$ & $0$ \\ \\ 
$0$ & $0$ & sin$^{2}\theta Y_{lm}$%
\end{tabular}
\right) \ ,
\end{equation}
and
\begin{equation}
\left( \hat{f}_{4}\right)_{ij}=\left( 
\begin{tabular}{ccc}
$0$ & $0$ & $0$ \\ \\
	$0$ & $\partial_{\theta }^{2}Y_{lm}$ & 
	$\left( \partial_{\theta \phi}^{2}-\cot \theta 
	\partial_{\phi }\right) Y_{lm}$ \\ \\
$0$ & $\left( \partial_{\theta \phi }^{2}-\cot \theta 
	\partial_{\phi }\right) Y_{lm}$ & $\left( \partial_{\phi }^{2}
	-\sin \theta \cos \theta \partial_{\theta }\right) Y_{lm}$%
\end{tabular}
\right) . 
\end{equation}

	Also for even-parity perturbations, the gauge freedom can be
exploited and in particular a gauge can be chosen in which
\begin{equation}
G=h_{0}=h_{1}=0 \ .
\end{equation}
As a result of this gauge choice, the polar metric perturbations assume
the more compact form
\begin{equation}
\label{hmn_pol}
h^{\rm pol}_{\mu \nu}=\left( 
\begin{tabular}{cccc}
$H_{0}\left( 1-2M/r\right) $ & $H_{1}$ & $0$ & $0$ \\ \\
$H_{1}$ & $H_{2}\left( 1-2M/r\right)^{-1}$ & $0$ & $0$ \\ \\ 
$0$ & $0$ & $r^{2}K$ & $0$ \\ \\
$0$ & $0$ & $0$ & $r^{2}\sin^{2}\theta K$%
\end{tabular}
\right) P_{l}\left( \cos \theta \right) e^{im \phi}\ .
\end{equation}
	Writing now out the Einstein equations for the perturbed metric
(\ref{hmn_pol}) leads to the following equation
\begin{equation}
\label{zer}
\frac{\partial^{2}Z}{\partial t^{2}}-
	\frac{\partial^{2}Z}{\partial r_*^{2}}+\widetilde{V}Z=0 \ ,
\end{equation}
which is also known as the ``Zerilli equation''. The explicit form of the
``Zerilli function'' is rather involved but can be expressed,
independently of the gauge chosen, as
\begin{equation}
Z \equiv \frac{4 r e^{-4\lambda} k_2 + l(l+1)r k_1}{l(l+1) - 2 + 6M/r}\ ,
\end{equation}
where $e^{-\lambda}\equiv 1 - 2M/r$ and the functions $k_1, k_2, k_3,
k_4$ are introduced in place of $G, h_1, K, H_2$ and are defined through
the relations
\begin{eqnarray}
G   &=&  k_3 \ , 
\\ \nonumber \\ 
h_1 &=&  k_4 \ , 
\\ \nonumber \\ 
K   &=&  k_1 - \frac{e^{-2\lambda}}{r}\left[r^2 
	\frac{\partial k_{3}}{\partial r} - 2 k_4\right]  \ , 
\\ \nonumber \\ 
H_2 &=&  2 e^{-2\lambda} k_2 + r \frac{\partial k_{1}}{\partial r} + 
	\left(1 + r\frac{\partial \lambda}{\partial r}\right)k_1  
	- e^{-\lambda}\frac{\partial}{\partial r}\left[r^2 e^{-\lambda}
	\frac{\partial k_{3}}{\partial r} - 
	2 e^{-\lambda}k_4\right]  \ .
\end{eqnarray}

	Note that as for odd-parity perturbations, the Einstein equations
can be recast in the form of a wave equation in a scattering potential
barrier ${\widetilde V}$, defined as
\begin{equation}
{\widetilde V}\equiv\left( 1-\frac{2M}{r}\right) \left[ \frac{2q\left( q+1\right)
	r^{3}+6q^{2}Mr^{2}+18qM^{2}r+18M^{3}}{r^{3}
	\left( qr+3M\right)^{2}}\right] \ ,
\end{equation}
where $q\equiv \left( l-1\right) \left( l+2\right)/2$. 

	Interestingly, the Regge-Wheeler an Zerilli equations (\ref{rw1})
and (\ref{zer}) are closely related and it is possible to transform the
first one for axial modes into the second one for polar modes via
suitable differential operators~\cite{c92}.

\subsection{QNMs of Black Holes}

	Since equations (\ref{rw1}) and (\ref{zer}) describe the response
of the black hole to external perturbations, they are basically telling
us about the {\it vibrational modes} of such a spacetime. In particular,
if a harmonic time dependence is introduced for the perturbations in
equations (\ref{rw1}) and (\ref{zer}), i.e. if $Q, Z \sim \exp (i
\omega_n t)$ where $\omega_n$ is the oscillation frequency of the $n$-th
mode and is a complex number of the type
\begin{equation}
\omega_n = \omega_{r,n} + i \omega_{i,n}\ , \qquad  \qquad 
	{\rm with}\ \  n = 0, 1, 2,\ldots \ , 
\end{equation}
it is then possible to define the {\it Quasi-Normal Modes} (QNMs) of the
black hole as the solutions of equations 
\begin{eqnarray}
&& \partial_{r_*}^{2}Q+\left[ \omega^{2}-        V \right] Q=0 
	\ ,\\ \nonumber \\ 
&& \partial_{r_*}^{2}Z+\left[ \omega^{2}-{\widetilde V}\right] Z=0 \ ,
\end{eqnarray}
that satisfy a {\sl pure outgoing-wave} boundary condition at spatial
infinity and a {\sl pure ingoing-wave} boundary condition at the event
horizon, i.e.
\begin{equation}
Q, Z \sim \exp(i  \omega r_*) \quad {\rm for}\ \ r_*\rightarrow -\infty\ ,
	\qquad {\rm and} \qquad
Q, Z \sim \exp(-i \omega r_*) \quad {\rm for}\ \ r_*\rightarrow \infty \ .
\end{equation}

\subsubsection{A Summary of Main Results}

	The literature on the solution of the Regge-Wheeler and of the
Zerilli equations as well as on the determination of the perturbation
spectrum of black holes is vast. The interested reader will find detailed
discussions on these topics and the relevant references in the review by
Kokkotas and Schmidt~\cite{ks99} and in the one by Ferrari~\cite{f01}. In
what follows, however, I briefly summarize what could be considered the
main results in the solution of the eigenvalue problem for the QNMs of a
Schwarzschild black hole:

\begin{itemize}

\item All the QNMs of Schwarzschild black hole have positive imaginary
parts and represent therefore damped modes. As a result, a Schwarzschild
black hole is {\sl linearly stable} against perturbations.

\item The damping time of these perturbations depends linearly on the
mass of the black hole (i.e. $\omega_{n} \sim 1/ M$) and is shorter for
higher-order modes (i.e. $\omega_{i,n+1} > \omega_{i,n}$). As a result,
the detection of gravitational waves emitted from a perturbed black hole
could provide a direct measure of its mass.

\item The excitation of a black hole and the consequent emission of
gravitational radiation is referred to as black hole ``ringing''. The
amplitudes of the gravitational waves emitted during the black hole
ringing decay in time and the late-time behaviour (or tail of the
ringing) is such that the amplitude evolution can be described in terms
of a power law, whose envelope represents the superposition of the
various QNMs.

\item The QNMs in black holes are {\it isospectral}, i.e. axial {\it and}
polar perturbations have the same complex eigenfrequencies so that the
real and imaginary parts of the spectrum are identical. This is simply
due to the uniqueness in which a black hole can react to a
perturbation. This is not true for relativistic stars.

\item The fundamental frequencies of oscillations $\omega_{r,n}/2 \pi$
have been computed by a number of authors and are now known up to very
large mode numbers. Reported in the table below are the frequencies of
the first four modes, together with the values of their decaying times
$\tau_{\rm d}$ (i.e. $1/\omega_{i,n}$). The data refers to modes with
$l=2,3$ and have been computed for $M=M_\odot$.
\begin{equation*}
\begin{tabular}{||c|c|c|c||c|c|c|c||}
\hline 
$l$ &$n$ & $\omega_{r,n}/2 \pi$ (kHz) & $\tau_{\rm d}$ (ms) & 
$l$ &$n$ & $\omega_{r,n}/2 \pi$ (kHz) & $\tau_{\rm d}$ (ms) \\ 
\hline 
  &    &          &          &        &          & &	    \\
2 &$0$ & $12.075$ & $5.5344\times 10^{-2}$ & 
3 &$0$ & $19.376$ & $5.3135\times 10^{-2}$ \\ 
2 &$1$ & $11.203$ & $1.7983\times 10^{-2}$ & 
3 &$1$ & $18.833$ & $1.7510\times 10^{-2}$ \\ 
2 &$2$ & $9.7291$ & $1.0298\times 10^{-2}$ & 
3 &$2$ & $17.834$ & $1.0281\times 10^{-2}$ \\
2 &$3$ & $8.1264$ & $6.9856\times 10^{-3}$ & 
3 &$3$ & $16.551$ & $7.1354\times 10^{-3}$ \\
  &    &          &          &        &          & &	    \\

\hline 
\end{tabular}
\end{equation*}

\item For any value of the harmonic index $l$, the real part of the
frequency $\omega_{r,n}(l)$ approaches a nonzero limiting value as the
mode number $n$ increases, while the imaginary part increases linearly as
$\sim n/4$ (i.e. higher modes have shorter decaying timescales).

\item Most of what has been presented in this Section for a Schwarzschild
black hole can be formulated also for a rotating (or Kerr) black hole. In
this case, however, the mathematical apparatus is more involved (the
potential is, for instance, complex) and some new features, such as the
{\it super-radiance} (i.e. the amplified scattering of electromagnetic
waves) can take place~\cite{sc73,pt73}. More on this can be found
in~\cite{ks99,f01}

\end{itemize}

\newpage
\section{Gravitational Waves from Perturbed Stars}

	Black holes are often considered the most ``extreme'' objects
predicted by General Relativity in the sense of being the objects with
the most intense gravitational fields. This is of course correct, but it
is worth reminding that relativistic stars (such as neutron stars or
strange stars) are comparable to black holes in this respect and
certainly the most relativistic ``astrophysical'' objects in the
Universe. An important way in which they differ from black holes,
however, is in possessing a surface and an interior structure which is
not causally disconnected from the exterior spacetime. A rough estimate
of how relativistic (and therefore ``extreme'') a compact star can be, is
provided by its {\it compactness}, i.e. by the ratio between its
gravitational mass $M$ and its radius $R$. Shown in the table below is
this quantity for a number of different objects
\medskip
\begin{equation*}
\begin{tabular}{||l|c||}
\hline 
 &$M/R$ \\
\hline 
solar-type star 	& $\sim 10^{-6}$ 	\\
white dwarf 		& $\sim 10^{-4}$ 	\\
neutron star 		& $0.1-0.3$ 		\\
black hole 		& $0.5$ 		\\
\hline
\end{tabular}
\end{equation*}
\medskip

\begin{figure}[htb]
\begin{center}
\hspace{0.125truecm}
\leavevmode

\hbox{\psfig{figure=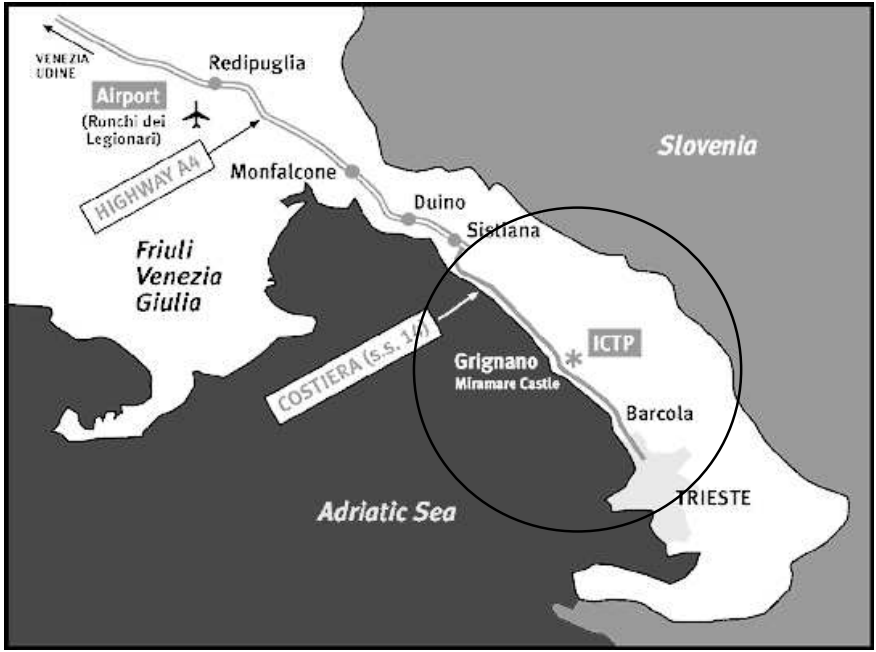,width=14.0truecm,angle=-90} }
\caption{\footnotesize Schematic representation of the spatial dimensions
of a neutron star whose surface is shown as an open circle and is
superposed over a local map of town for comparison. (This figure has been
elaborated from one available on the AS-ICTP website.)
\label{fig4}}
\end{center}
\end{figure}

	Clearly, while an ordinary star cannot be considered a compact
object, the table above shows that just a small difference in compactness
distinguishes a neutron star from a black hole and this justifies the
interest that relativists have reserved to this type of objects. In order
to have a more concrete example of how ``compact'' a relativistic star
can be, I have shown in Figure~\ref{fig4} a schematic representation of
the spatial dimensions of a neutron star superposing its surface to a
local map of town for comparison. One should therefore imagine that an
amount of matter a million times larger than the one contained in the
Earth can be concentrated in an object with a radius of just ten
kilometres.

	Just like black holes, compact stars can respond to perturbations
and emit gravitational waves. Quite differently from black holes,
however, the gravitational radiation coming from perturbed relativistic
stars is extremely rich of details, many of which carry important
physical information. Imprinted in these gravitational waves, in fact, is
a detailed map of the internal structure of the emitting stars, which can
be used to deduce the properties of matter at conditions that cannot be
investigated by experiments in terrestrial laboratories.

	Despite these exciting perspectives, the issue of the {\it
detectability} of the gravitational radiation from perturbed relativistic
stars is still basically unsettled. This is largely due to our ignorance
about the precise physical conditions leading to a perturbed relativistic
star. A simple example in this sense is offered by a protoneutron star
formed after the gravitational collapse in a supernova explosion. While
it is generally expected that the newly born neutron star will pulsate
wildly during the first few seconds following the collapse, how much
energy will be transferred to the pulsation and subsequently radiated
through the oscillation modes is unknown. The only realistic way of
overcoming this ignorance is to perform detailed, fully relativistic
calculations and deduce from them how large the perturbations will
be. While the recently developed numerical codes will soon be able to
provide some quantitative answer in this respect~\cite{fetal00, su00,
fetal02, detal02, su02, duez_etal02}, at present one can simply argue
that {\it a)} the energy stored in the pulsation can potentially be of
the same order as the kinetic energy of the system; {\it b)} the
oscillations will be damped mainly through the emission of gravitational
waves so that the released energy could be considerable. 

	Under these assumptions, the effective gravitational wave
amplitude $h$ for a star oscillating in its fundamental mode of
oscillation can be estimated simply. For weak gravitational waves, in
fact, the gravitational wave luminosity (i.e. the rate of energy loss to
gravitational waves) at a distance $d$, can be written to be~\cite{s96}
\begin{equation}
\label{edot}
\frac{dE}{dt} \simeq \frac{E}{\tau}
	\simeq 4 \pi d^2 \left( \frac{c^3}{16\pi G} \right)
	|{\dot h}|^2 \ ,
\end{equation}
where $E$ is the total energy lost over the time $\tau$. If the star is
oscillating at a frequency $f$, then ${\dot h} \approx 2 \pi f h$ and
expression (\ref{edot}) can be rewritten as
\begin{equation}
\label{h}
h \simeq 1.2\times 10^{-21} 
	\left( \frac{E}{\widetilde E}  	\right)^{1/2} 
	\left( \frac{1\ {\rm ms}}{\tau} \right)^{1/2} 
	\left( \frac{1\ {\rm kHz}}{f}  	\right)
	\left( \frac{50\ {\rm kpc}}{d} 	\right) \ ,
\end{equation}
where ${\widetilde E}=8.2\times 10^{-8}\ M_\odot c^2$ is the energy lost
in gravitational waves as estimated through recent relativistic
calculations~\cite{detal02}. Note that the probability of detecting a
source can be increased if suitable data analysis techniques, such as
``matched filtering'', are used~\cite{ct01,h02}. In this case, it is
possible to estimate the ``effective'' gravitational wave amplitude
$h_{\rm eff}$ to be $h_{\rm eff} \simeq h \sqrt{f \tau}$, so that
(\ref{h}) becomes
\begin{equation}
\label{heff}
h_{\rm eff} \simeq 1.2\times 10^{-21} 
	\left( \frac{E}{\widetilde E}  	\right)^{1/2} 
	\left( \frac{1\ {\rm kHz}}{f}  	\right)^{1/2} 
	\left( \frac{50\ {\rm kpc}}{d} 	\right) \ .
\end{equation}

	The distance scale $d$ used in expressions (\ref{h}) and
(\ref{heff}) is that to the supernova SN1987A and the number of events in
the corresponding volume is one every 10--20 years.  Clearly, this event
rate is too small for being of interest and it is therefore necessary to
consider a volume much larger, such as the one comprising the Virgo
cluster, to reach an event rate of a few per year. In this case, it is
possible to consider the problem of the detection from a different point
of view and rather calculate what is the energy $E$ necessary to obtain
an effective wave amplitude $h_{\rm eff} \sim 10^{-21}$ for a source at a
distance $d = 20$ Mpc. Using expression (\ref{heff}), the answer is $E
\approx 0.01 \ M_\odot$. While these estimates may appear optimistic,
they cannot be ruled out and the pay-offs of a potential detection would
be so great to justify the intense research this field is experiencing.

	To better appreciate how the detection of gravitational waves
from pulsating relativistic stars can be used to derive information about
the physical properties of the source it is necessary to consider in more
detail the perturbation theory of fluid stars, both in relativistic and
in Newtonian gravity. This is the scope of the following Sections, where
I will first discuss the spherical oscillations of relativistic stars and
then the nonradial oscillations of Newtonian stars. A brief
classification of the oscillation modes and an introduction to the onset
of non-axisymmetric instabilities in rotating star will conclude these
lecture notes.

\subsection{Linear Perturbations of Fluid Stars}

	Fluid stars have two fundamental classes of oscillation modes
that are referred to as {\it nonradial} and {\it radial} according to
whether the perturbed quantities have an angular dependence or
not. Furthermore, since not constrained by Birkhoff's theorem
(cf. Section \ref{pobhs}), nonrotating (or spherical) stars are allowed
to also have radial oscillations in addition to the nonradial ones. In
the following sections I will discuss both radial and nonradial
oscillations of nonrotating stellar models, treating the first ones in
full General Relativity, while restricting the discussion to Newtonian
gravity for the second ones.

	As done for the perturbations of black holes, I first consider a
fluid star which has a well defined ``unperturbed'' state which is one of
equilibrium. The star will then be considered in a new ``perturbed''
state, not necessarily of equilibrium, which represents a small deviation
from the initial one, i.e. 
\begin{equation}
f(t,r,\theta,\phi) = f_{0}(r) + f_{1}(t, r,\theta,\phi) \ ,
\end{equation}
where $f_{0}(r)$ is the value of the unperturbed quantity (considered
here to be spherically symmetric) and $f_{1}(t, r,\theta,\phi)$ is its
perturbation (considered here to be have also an angular dependence). Of
course, linearity is guaranteed by the condition that $f_{1}/f_{0} \ll
1$.

	Two different approaches to the description of infinitesimal
perturbations are possible and in which the physical variables are seen
as

\begin{itemize}
	
\item {\it Eulerian perturbations:} when the perturbed variables are
considered at a fixed position in space with coordinates ${\vec x}$, i.e.
\begin{equation}
\delta f(t, {\vec x}) \equiv f(t, {\vec x}) - 
	f_{0}({\vec x})\ ,
\end{equation}
where $f(t, {\vec x})$ is the new value assumed by the physical variable
(e.g. $\rho$, $p$) and $f_{0}({\vec x})$ is the equilibrium value,
independent of time.

\item {\it Lagrangian perturbations:} when the perturbation is obtained
after comparing variables at two different spatial points occupied by the
same fluid element
\begin{equation}
\Delta f(t, {\vec x}) \equiv f(t, {\vec x} + 
	{\vec {\xi}}(t, {\vec x})) - 
	f_{0}({\vec x}_0)\ ,
\end{equation}
where ${\vec \xi}(t, {\vec x})\equiv {\vec x} - {\vec x}_0$ is the
Lagrangian displacement three-vector (${\vec \xi}=0$ in the unperturbed
configuration), $f(t, {\vec x} + {\vec \xi}(t, {\vec x}))$ is the value
of the perturbed physical variable, and $f_0 ({\vec x}_0)$ the
equilibrium value in the initial position ${\vec x}_0$.

\end{itemize}
A direct relation between the two descriptions is expressed through the
following relation between the two operators $\Delta$ and $\delta$
\begin{equation}
\label{eul_lag}
\Delta  = \delta   + {\cal L}_{\vec \xi}  \ ,
\end{equation}
where ${\cal L}_{\vec \xi}$ is the Lie derivative along $\vec \xi$.

\subsection{Radial Oscillations: Relativistic Stars}

	Consider the unperturbed star to be composed of a perfect fluid
with four-velocity ${\boldsymbol {\it u}}_0$, energy density $\rho_0$ and isotropic
pressure $p$, whose stress-energy tensor takes the form
\begin{equation}
\Tzdmn =(\rho_{0}+p_{0})
	(u_0)_{\mu}(u_0)_{\nu} + p_0 \gzdmn \ ,
\end{equation}
where 
\begin{equation}
n_0 \equiv \frac{\rho_*}{m_0} \ ,
\end{equation}
is the baryon number density of fluid particles of rest-mass $m_0$ and
rest-mass density $\rho_*$, so that $\epsilon_0=\rho_0/\rho_* - 1$ is the
unperturbed internal energy density. Note that, as in equation
(\ref{dsz}), I have here used the upper zero index to indicate tensors
referring to unperturbed quantities. The generic background spacetime of
such a static spherical star is expressed through the line element
\begin{equation}
\label{dsz_tov}
ds^{2}=-e^{2 \nu_{0}(r)} dt^{2} + e^{2 \lambda_{0}(r)} dr^{2} + 
	r^{2}(d\theta^{2}+\sin^{2}{\theta}d\phi^{2}) \ ,
\end{equation}
so that the Einstein equation in such a spacetime can be recast as a set
of ordinary differential equations
\begin{eqnarray}
\label{tov1}
&&\frac{dp_0}{dr} = - \frac{\rho_0 m}{r^2}\left(1 + \frac{p_0}{\rho_0}\right)
	\left(1 + \frac{4\pi p_0 r^3}{m}\right)\left(1 - 
	\frac{2m}{r}\right)^{-1}\ ,
\\ \nonumber \\
&&\frac{d\n_0}{dr} = - \frac{1}{\rho_0} \frac{dp_0}{dr} 
	\left(1 + \frac{p_0}{\rho_0}\right)^{-1}\ ,
\\ \nonumber \\
\label{tov3}
&&\frac{dm}{dr} = 4 \pi r^2 \rho_0 \ ,
\end{eqnarray}
where the function $m=m(r)$ has been introduced to re-express the
$g_{rr}$ metric function, i.e.
\begin{equation}
e^{2\lambda_0} \equiv \left(1 -\frac{2m}{r}\right)^{-1} \ .  
\end{equation}
The function $m(r)$ has the interpretation of gravitational ``mass inside
the radius $r$'' because the integral form of (\ref{tov3}) can be written
as
\begin{equation}
M \equiv \int^R_0 4 \pi r^2 \rho_0 dr \ ,
\end{equation}
where $R$ is the radius of the star, so that $M$ represents the total
mass-energy of the star.

	The system of equations (\ref{tov1})--(\ref{tov3}) in the
variables $p_0, \n_0, \rho_0,$ and $m$ is usually referred to as the set
of Tolmann-Oppenheimer-Volkoff (or TOV) equations~\cite{t39,ov39}. An
additional equation is necessary to close the system and this is given by
an equation relating the energy density and pressure appearing in the
stress-energy tensor. Such a relation is usually referred to as {\it
equation of state} (EOS) and in the case of a fluid in local
thermodynamic equilibrium there always exists a relation of the type
\begin{equation}
p_0 = p_0 (\rho_0, s_0) \ ,
\end{equation}
where $s_0$ is the specific entropy. If the change in the fluid entropy
are very small so that the latter can be considered a small constant
quantity, the EOS can be simply written as a relation between the energy
density and the pressure and is referred to as {\it barotropic}. A very
common example of a barotropic EOS is a {\sl polytropic} EOS
\begin{equation}
p_0 = K \rho^{\Gamma}_0 \ ,
\end{equation}
where $K$ and $\Gamma$ are the polytropic constant and exponent,
respectively. 

	Once the ``background'' static stellar model has been constructed
as a solution of the TOV equations, a radial perturbation of Eulerian
type can be introduced in both the metric variables
\begin{eqnarray}
\label{r-perts-1}
\nu_0(r) & \longrightarrow & \nu(t,r)=\nu_{0}(r)+\delta \nu(t,r) \ , 
\\ \nonumber \\
\label{r-perts-2}
\lambda_0(r) & \longrightarrow & \lambda(t,r)=\lambda_{0}(r)+
	\delta \lambda(t,r) \ , 
\end{eqnarray}
and in the fluid variables
\begin{eqnarray}
\label{r-perts-3}
p_0(r) & \longrightarrow & p(t,r)=p_{0}(r)+\delta p(t,r) \ , 
\\ \nonumber \\
\label{r-perts-4}
\rho_0(r) & \longrightarrow & \rho(t,r)=\rho_{0}(r)+\delta \rho(t,r) \ , 
\\ \nonumber \\
\label{r-perts-5}
n_0(r) & \longrightarrow & n(t,r)=n_{0}(r)+\delta n(t,r) \ .
\end{eqnarray}
where I have assumed the perturbations to be isentropic, so that $\delta
K=0=\delta \Gamma$.

	Note that as a result of (\ref{r-perts-1})--(\ref{r-perts-2}),
the metric line element (\ref{dsz_tov}) takes the non-diagonal form
\begin{equation}
ds^{2}=-e^{2 \nu(r^{\prime},\;t)} dt^{2} + 2Adr^{\prime}dt + 
	e^{2 \lambda(r^{\prime},\;t)}dr^{\prime 2} + 
	r^{\prime 2}(d\theta^{2}+\sin^{2}\theta d\phi^{2}) \ ,
\end{equation}
which, however, can always be recast in a diagonal form of the type 
\begin{equation}
ds^{2}=-e^{2 \nu(t,r)} dt^{2} + e^{2 \lambda(t,r)}dr^{2} + 
	r^{2}(d\theta^{2}+\sin^{2}\theta d\phi^{2}) \ ,
\end{equation}
after an appropriate coordinate transformation.

	Because only motions in the radial direction are possible, the
problem has only one degree of freedom and it is convenient to use the
radial component of the Lagrangian displacement vector
$\xi^{i}=(\xi(t,r),0,0)$ to express all of the equations in terms of this
scalar function. In particular, since the perturbed Eulerian velocity has
components $u^{r}=0+\delta u^{r}$ and $u^{t}=u_{0}^{t}+\delta u^{t}$, the
usual normalization conditions
\begin{equation}
u^{\mu}u_{\mu}=-1\ , \qquad 
	(u_{0})^{\m} (u_{0})_{\mu}=-1 \ ,
\end{equation} 
together with the relation
\begin{equation}
\Delta u^a = \frac{1}{2}u^a u^b u^c \Delta g_{bc} \ ,
\end{equation} 
can be used to derive the following expressions for the perturbed
components
\begin{eqnarray}
\label{pert_4v}
& & \delta u^{r} = e^{-\nu_{0}} \dot{\xi} \ ,
\\ \nonumber \\
& & \delta u^{t} = e^{-\nu_{0}} \delta \nu \ .
\end{eqnarray}
Expression (\ref{pert_4v}) shows that, at first order in the
perturbations, the time derivative of the Lagrangian radial displacement
$\dot{\xi}$ is related to the radial component of the three-velocity,
i.e.
\begin{equation}
\frac{u^{r}}{u^{t}} = \frac{\delta u^r}{u^t_0} =
	\frac{dr}{dt} = {\dot \xi} \ ,
\end{equation}
where $dr/dt$ is calculated along the worldline of the unperturbed
star. The set of perturbative equations to be solved is therefore 
\begin{eqnarray}
\label{deq1}
&\delta\left[G_{\mu\nu}-8 \pi T_{\mu\nu}\right]=0  \ & 
	\qquad {\rm Einstein\ equations,}
\\ \nonumber \\
\label{deq2}
&\delta\left[u^{\mu}\nabla_{\alpha}T^{\alpha}_{\ \mu}\right] = 0 \ & 
	\qquad {\rm energy\ conservation,}
\\ \nonumber \\
\label{deq3}
&\delta\left[ P^{\mu}_{\ r} \nabla_{\alpha}T^{\alpha}_{\ \mu}\right]=0 \ & 
	\qquad	{\rm momentum\ conservation,}
\\ \nonumber \\
\label{deq4}
&\delta\left[\nabla_{\mu}(n u^{\mu})=0\right] \ &
	 \qquad {\rm baryon\ number\ conservation,}
\\ \nonumber \\
\label{deq5}
&\displaystyle \frac{\Delta p}{p}=\Gamma_{1}\frac{\Delta n}{n} \ & 
	\qquad {\rm adiabatic\ condition,}
\end{eqnarray}
where $P^{\mu}_{\ \nu}=(u_0)^{\mu} (u_0)_{\nu} + \gzumdn$ is the
projection tensor orthogonal to ${\boldsymbol {\it u}}_0$ and
\begin{equation}
\Gamma_{1} = \Gamma_{1}(r) \equiv 
	\frac{\partial \ln p}{\partial \ln \rho}\Bigg|_s 
	\ ,
\end{equation}
and is assumed to be a constant in time.

	The full set of perturbative equations can be derived after
expanding equations (\ref{deq1})--(\ref{deq4}) and retaining only the
first-order terms. In particular, the baryon conservation equation
(\ref{deq4}) has a perturbed expression given by
\begin{equation}
\label{deqn1}
\displaystyle \Delta n = -n_{0}\left[
	\frac{1}{r^2 e^{\lambda_{0}}}
	(r^{2}e^{\lambda_{0}}\xi)^{\prime}+
	\delta \lambda \right] \ ,
\end{equation}
with the prime indicating a partial radial derivative. Similarly, the
perturbed energy and adiabatic conservation equations (\ref{deq2}) and
(\ref{deq5}) yield the following relation between the perturbed energy
density and baryon number
\begin{equation}
\label{deqn2}
\Delta \rho = \frac{\rho_{0}+p_{0}}{n_{0}}\Delta n \ .
\end{equation}
Equation (\ref{deqn2}) can also be expressed in terms of Eulerian
perturbations to give
\begin{eqnarray}
\label{deqn2_1}
&& \delta \rho =
	-(\rho_{0}+p_{0})\left[\frac{1}{r^2 e^{\lambda_{0}}}
	(r^{2}e^{\lambda_{0}}\xi)^{\prime}+\delta
	\lambda\right]-\xi\rho^{\prime}_{0} \ ,
	\\ \nonumber \\
&& \delta p = -\Gamma_{1} p_{0}\left[
	\frac{1}{r^2 e^{\lambda_{0}}}
	(r^{2}e^{\lambda_{0}}\xi)^{\prime}+\delta \lambda 
	\right]-\xi p^{\prime}_{0} \ .
\end{eqnarray}
Next, the perturbations in the metric functions described by the
field equations (\ref{deq1}) yield
\begin{eqnarray}
&& \delta \lambda = -4 \pi (\rho_{0}+p_{0}) r
	e^{2 \lambda_{0}}\xi \ ,
\\ \nonumber \\
\label{deqn2_4}
&& \delta \nu^{\prime}=- \frac{4 \pi \Gamma_{1}p_{0}}{r} e^{2 \lambda_{0}
	+ \nu{0}}\left(\frac{r^{2}}{e^{\nu_{0}}}\xi\right)^{\prime}+ [4
	\pi p^{\prime}_{0}r - 4 \pi (\rho_{0}+p_{0})]e^{2 \lambda_{0}}\xi
	\ .
\end{eqnarray} 
Finally, the perturbed Euler equation (\ref{deq3}) yields the perturbed
``equation of motion''

\begin{equation}
\label{deqn4}
(\rho_{0}+p_{0})e^{2(\lambda_{0} - \nu_{0})}\ddot{\xi}=-\delta
	p^{\prime}-(\delta \rho + \delta p)
	\nu_{0}^{\prime}-(\rho_{0}+p_{0})\delta \nu^{\prime} \ ,
\end{equation}
whose right-hand-side shows the well-know restoring forces due to
gradients in the pressure or in the gravitational potentials (both the
unperturbed and the perturbed one). 

	It is now convenient to introduce the following auxiliary
variables
\begin{eqnarray}
{\widetilde \eta} &\equiv& r^{2} e^{-\nu_{0}}\xi\ ,
\\
W &\equiv& (\rho_{0} + p_{0}) \frac{e^{3\lambda_{0}+
	\nu_{0}}}{r^{2}} > 0\ ,
\\
A &\equiv& \Gamma_{1} p_{0} \frac{e^{\lambda_{0}+
	3\nu_{0}}}{r^{2}} > 0 \ ,
\\
Q &\equiv& e^{\lambda_{0}+3\nu_{0}}
	\bigg[\frac{(p_{0}^{\prime})^{2}}{(\rho_{0}+p_{0})r^2} - 
	\frac{4p_{0}^{\prime}}{r^{3}} - 8 \pi (\rho_{0}+p_{0})p_{0}
	\frac{e^{2\lambda_0}}{r^2}\bigg] \ ,
\end{eqnarray}
in terms of which equation (\ref{deqn4}) can be written [after using
eqs.~(\ref{deqn2_1})--(\ref{deqn2_4})] as an inhomogeneous wave equation
of the type
\begin{equation}
\label{iwe}
W\ddot{\widetilde \eta}=(A {\widetilde \eta}^{\;\prime})^{\prime} + 
	Q{\widetilde \eta} \ .
\end{equation}
As done in Section~\ref{pobhs} for the perturbations of a black hole, a
harmonic time dependence of the perturbations can now be assumed in terms
of a constant complex frequency $\sigma$
\begin{equation}
{\widetilde \eta}(t,r)=\eta(r)e^{-i \sigma t}\ ,
\end{equation}
so that equation (\ref{iwe}) becomes
\begin{equation}
\label{sle}
(A\eta^{\prime})^{\prime}+(Q+\sigma^{2}W)\eta=0 \ .
\end{equation}
Equation (\ref{sle}) represents a typical {\it Sturm-Liouville}
eigenvalue equation and, together with suitable {\it boundary
conditions}, its solution provides the eigenvalues and the eigenfunctions
for the radial perturbations. Fortunately, for spherical stars the
boundary conditions at the extrema of the range where the eigenvalue
problem needs to be solved are rather straightforward to formulate and
can be summarized as follows:

\begin{itemize}

\item when $r \rightarrow 0$, all quantities should be regular
(i.e. either zero or finite) and this then implies that $\xi$ and
$\xi^{\prime}$ should be regular or, equivalently, that $\eta\sim r^3$
for $r \rightarrow 0$.

\item when $r \rightarrow R$, on the other hand, the perturbation should
leave the Lagrangian pressure perturbation equal to zero at the stellar
surface, i.e. $\Delta p=0$ for $r \rightarrow R$. In practice, then, this
implies that
\begin{equation}
-\frac{e^{\nu_0}}{r^2} {\Gamma_1 p_0}\eta^{\prime} = 0 \qquad 
	{\rm for}\ r \rightarrow R \ .
\end{equation} 
\end{itemize}

	Despite the fact that determining the eigenfunctions and
eigenfrequencies of radial oscillations represents the simplest possible
eigenvalue problem, doing this in General Relativity requires the
numerical solution of equation (\ref{sle}). I will not discuss here the
details of how this is done in practice, but simply summarize in what
follows the main results~\cite{btm66,mtw74,gl83}:
\begin{itemize}

\item The eigenvalues $\sigma_n$ form an infinite discrete sequence with
integer index $n$, i.e.

\begin{equation}
\sigma_{0}^{2}<\sigma_{1}^{2}<\sigma_{2}^{2}< \ldots \ .
\end{equation}

\item The eigenvalues $\sigma^{2}$ are all real.

\item The eigenfunction corresponding to the lowest eigenfrequency
$\eta_0(r)$ has zero nodes for $0 < r < R$, while the $n$-th eigenfuction
$\eta_n(r)$ has $n$  nodes in the same interval.

\item The eigenfunctions $\eta_n(r)$ are orthonormal with weight function
$w(r)$, i.e.

\begin{equation}
\int^R_0 \eta_n \eta_m w(r) dr = \delta_{n m} \ .
\end{equation}

\item When using a description of the Sturm-Liouville problem in terms
of a variational principle, the eigenvalues can then be calculated as
\begin{equation}
\label{vp}
\sigma^{2}={\rm extremal~of~} \left[ 
	\frac{ \int_0^R \left[A( \eta^{\prime})^2 - Q \eta^2\right] dr}
	{\int_0^R w\eta^2 dr} \right] \ .
\end{equation}
While formally elegant, the variational method is not very used in
numerical calculations~\cite{btm66}.

\item The absolute minimum of (\ref{vp}) signs the frequency of the
fundamental mode of pulsation. If this quantity is negative, the mode
will grow in time on a timescale $\tau_{\rm g}$ [i.e. $\eta \sim
\exp(-i\sigma t) \sim \exp(t/\tau_{\rm g})$] and the star will be {\it
unstable} to radial oscillations. If, on the other hand, this quantity is
positive, the mode will decay on a timescale $\tau_{\rm d}$ [i.e. $\eta
\sim \exp(-i\sigma t) \sim \exp(-t/\tau_{\rm d})$] and the star will be
{\it stable}. Since the denominator of (\ref{vp}) is positive-definite,
the stability against radial perturbations is given by
\begin{equation}
\label{sc}
{\rm radial\ stability} \quad \Longleftrightarrow \quad 
\sigma ^2 > 0 \quad \Longleftrightarrow \quad 
	\int_0^R[A(\eta^{\prime})^2 - Q\eta^2]dr > 0\ .
\end{equation}

\end{itemize}

	It is also very instructive to consider the solution of equation
(\ref{vp}) in the first-order post-Newtonian (1PN) approximation of
General Relativity, in which all of the variables are expanded up to
first order in the ratio $M/R$. In this case, the fundamental frequency
of oscillation $\sigma_0$ (not to be meant as an unperturbed quantity!)
can be written as
\begin{equation}
\label{pn1}
\sigma^2_0 = \frac{3 |W|}{I}({\bar \Gamma}_1-\Gamma_c)\ ,
\end{equation}
where $|W|$ is the star's gravitational binding energy, $I\equiv 4 \pi
\int (\rho_0 r^2) r^2 dr$ is the trace of the second moment of the mass
distribution and ${\bar \Gamma}_1$ is the pressure averaged adiabatic
index
\begin{equation}
\label{bg1}
{\bar \Gamma}_1 \equiv \frac{4\pi \int^R_0 \Gamma_1 p_0 r^2 dr}
	     {4\pi \int^R_0 p_0 r^2 dr} \ .
\end{equation}

	Equation (\ref{pn1}) can also be interpreted as a criterion
stating that the stability of the star to the fundamental mode of radial
oscillation depends on the values of the averaged adiabatic index ${\bar
\Gamma_1}$ when compared with the ``critical'' adiabatic index
\begin{equation}
\Gamma_c \equiv \frac{4}{3}\,+\,\alpha\frac{M}{R}\ ,
\end{equation}
where $\alpha > 0 $ is a constant defined in terms of the equilibrium
quantities~\cite{mtw74}
\begin{equation}
\label{alpha_c}
\alpha \equiv \frac{4 \pi}{3} \frac{R}{M |W|} \int^R_0
	\left( 
	3 \rho_0 \frac{m^2_0}{r^2} + 
	4 p_0 \frac{m_0}{r} \right)r^2 dr \ , 
\end{equation}
and $m_0(r)$ is the rest-mass inside the radius $r$. As a result, the
stability of a relativistic stellar model to radial perturbation can be
summarized as
\begin{eqnarray}
{\bar \Gamma}_1&\ge&\Gamma_c \qquad \Longrightarrow \qquad
	\mathrm{the~star~is}\ stable\ \mathrm{to~radial~perturbations},
	\nonumber\\ 
{\bar \Gamma}_1&<&\Gamma_c \qquad \Longrightarrow \qquad
	\mathrm{the~star~is}\ unstable\ \mathrm{to~radial~perturbations},
	\nonumber 
\end{eqnarray}

	Since for a Newtonian polytropic star the critical value of the
adiabatic exponent discriminating stability is $\Gamma_c|_{_{\mathrm
{Newt.}}} = 4/3$, equation (\ref{pn1}) expresses the fact that the 1PN
correction (and therefore the effects related to a stronger gravitational
field) are ${\cal O}(M/R)$, thus making the criteria of stability more
severe ($\Gamma_c|_{_{\mathrm{Newt.}}} < \Gamma_c|_{_{\mathrm{1PN}}}$).
Stellar models that are radiation-pressure dominated and very massive
white dwarfs are very well described by polytropes with $\Gamma_1=4/3$
and therefore the post-Newtonian contributions expressed in (\ref{pn1})
are very important to assess the stability of these objects against
radial perturbations.

\subsection{Nonradial Oscillations: Newtonian Stars}
\label{nr_ns}

	The discussion of the properties of nonradial oscillation modes
is sufficiently complex (already for nonrotating stellar models) that a
first approach is best done within a simpler scenario. For this reason,
hereafter the discussion will be made in terms of a Newtonian description
of physics. The interested reader will find a detailed treatment of this
subject in~\cite{uetal89} and a review on nonradial oscillations in
relativistic stars in~\cite{s02}.

	Consider therefore an equilibrium state of a nonrotating and zero
temperature star that can be described by the distribution of pressure,
density and gravitational potential, functions of the radial coordinate
$r$ only, i.e.
\begin{equation}
p_0=p_{0}(r) \ , \qquad \rho_0 =\rho_{0}(r) \ , \qquad
	\Phi_0 =\Phi_{0}(r) \ . 
\end{equation}
These functions are obtained as solutions of a system of differential
equations describing the equilibrium 
\begin{eqnarray}
\label{he}
&\displaystyle -\frac{1}{\rho}\frac{dp}{dr}-\frac{d\Phi}{dr} = 0 \ ,
\\ \nonumber \\
\label{N_poisson}
&\nabla^2 \Phi = 4\pi G\rho \ , & 
\\ \nonumber \\
\label{eos}
&p = p(\rho)\ , & 
\end{eqnarray}
and in the case of a polytropic EOS, this system can be cast into a
second order differential equation, the Lane-Emden equation, for $p$ or
$\rho$~\cite{c39}. The solution of the system (\ref{he})--(\ref{eos}) is
possible after suitable boundary conditions are imposed at the centre
(where the regularity of all the quantities should be required) and at
the surface of the star (where the gravitational potential should be
matched to the vacuum solution behaving as $r^{-1}$). Introducing now the
Eulerian perturbations, the different variables can be expanded as
\begin{eqnarray}
\label{nr-perts-1}
&& p(t,r,\theta,\phi) = p_{0}(r) + \delta p (t,r,\theta,\phi)\ ,
\\ \nonumber \\ 
\label{nr-perts-2}
&& \rho(t,r,\theta,\phi) =\rho_{0}(r)+\delta \rho(t,r,\theta,\phi) \ , 
\\ \nonumber \\ 
\label{nr-perts-3}
&& {\vec v}(t,r,\theta,\phi) = {\vec v}_0(r) + \delta{\vec v} = 
	\delta{\vec v}(t,r,\theta,\phi) \ ,
\\ \nonumber \\ 
\label{nr-perts-4}
&& \Phi(t,r,\theta,\phi) =\Phi_{0}(r)+\delta \Phi(t,r,\theta,\phi) \ ,
\end{eqnarray}
\noindent
and it is useful to compare expressions
(\ref{nr-perts-1})--(\ref{nr-perts-4}) with the corresponding
relativistic expressions (\ref{r-perts-1})--(\ref{r-perts-5}) to
appreciate the nonradial aspect of these perturbations.

	Using the Lagrangian displacement vector ${\vec \xi}$, the
Lagrangian velocity perturbation of the stellar matter is~\cite{fs78}
\begin{equation}
\Delta v^i = \partial_t\xi^i\ ,
\end{equation}
and in the nonrotating stellar case ($v_0^i=0$) it reduces to
\begin{equation}
\delta v^i = \partial_t\xi^i \ .
\end{equation}
Note that here the Lagrangian displacement does no longer need to have a
radial component only [cf. equation (\ref{gen_xi})]. Introducing now the
perturbed quantities $\delta p, \delta \rho ,\delta \Phi$ and $\delta
v^i$ in equations (\ref{he})--(\ref{eos}) and in the equation for the
conservation of baryon number, one obtains
\begin{eqnarray}
\label{nr_perts_21}
& \displaystyle \delta \rho +\nabla (\rho_{0}{\vec \xi} )=0 \ , &
\\ \nonumber \\
\label{nr_perts_22}
& \displaystyle -\sigma^{2}{\vec \xi} +\frac{1}{\rho_{0}}\nabla (\delta p) + 
	\nabla (\delta \Phi) +\frac{\delta \rho }{\rho_{0}}
	\nabla (\Phi_{0})=0 \ , & 
\\ \nonumber \\
\label{nr_perts_23}
& \displaystyle \frac{1}{r^{2}}\partial_{r}
	[r^{2}\partial_{r}(\delta \Phi) ]+
	\nabla_{\bot }^{2}(\delta \Phi) =4\pi G\delta \rho \ , &
\\ \nonumber \\
\label{adiabatic_cons}
& \displaystyle \frac{\Delta p}{p}=\Gamma_{1}
	\frac{\Delta \rho }{\rho } \ . &
\end{eqnarray}
where $\nabla_{\bot}^{2}$ is the two-dimensional Laplacian operator in
spherical polar coordinates (i.e. a two-dimensional gradient operator on
the $r=\mbox{const.}$ 2-sphere)
\begin{equation}
\nabla^{2}_{\bot} \equiv \frac{1}{r^{2}\sin^{2}\theta }\left[ \sin \theta
	\partial_{\theta }(\sin \theta \partial_{\theta })+
	\partial_{\phi}^{2}\right] \ .
\end{equation}
Using now the relation between the Eulerian and Lagrangian perturbations
(\ref{eul_lag}), equation (\ref{adiabatic_cons}) can be rewritten as
\begin{equation}
\frac{\delta p+\xi ({d p_{0}}/{dr})}{p_0}=\Gamma_{1}
	\frac{\delta \rho +\xi ({d \rho_{0}}/{dr})}{\rho_0} \ ,
\end{equation}
or, alternatively, as
\begin{equation}
\frac{\delta \rho }{\rho }=
	\frac{1}{\Gamma_{1}}\frac{\delta p}{p}-A \xi \ ,
\end{equation}
with
\begin{equation}
\label{schwarz_dis}
A \equiv \frac{d\ln \rho_{0}}{dr}- \frac{1}{\Gamma_{1}}\frac{d\ln
	p_{0}}{dr} = \left(\frac{1}{\Gamma} - \frac{1}{\Gamma_{1}}\right)
	\frac{d\ln p_{0}}{dr} \ ,
\end{equation}
and with the second equality in (\ref{schwarz_dis}) applying for a
polytropic EOS only.

\noindent The quantity $A$ defined in equation (\ref{schwarz_dis}) is the
{\it Schwarzschild discriminant} and is used in discussing of the
convective instability in a star~\cite{t78}.  In particular, if $A > 0$
somewhere in the star (or equivalently if $\Gamma_1 > \Gamma$), the
matter there will be unstable to convective motions. The Schwarzschild
discriminant is therefore related to restoring buoyancy forces and is
important to determine the frequency of gravity waves in the radial
direction or, equivalently, the frequency at which a fluid element may
oscillate around its equilibrium position under the influence of gravity
and pressure gradients. Such a frequency is called the {\it
Brunt-V\"{a}is\"{a}l\"{a}} frequency and is defined as
\begin{equation}
\label{bv_f}
N^2 \equiv - A g \ ,
\end{equation}
where $g$ is the local gravitational acceleration
\begin{equation}
g\equiv \frac{GM(r)}{r^{2}} = \frac{1}{\rho_0}\frac{d p_0}{dr} \ .
\end{equation}

	A simplification should now be introduced to handle the
non-trivial angular dependence in the perturbations
(\ref{nr-perts-1})--(\ref{nr-perts-4}). To this scope I will assume that
all of the relevant perturbative equations can be rewritten after
decomposing the variables into vector spherical harmonics. In particular,
given a three-vector $\vec V$ with separable variables, it can be
decomposed into its ``polar'' and ``axial'' parts [cf. eq. (\ref{vsh})]
as
\begin{equation}
\label{dec_vh}
\vec V (r,\theta,\phi) = \sum_{l,m} [ V_{P1}(r) 
	Y_{lm}{\vec e}_r + V_{P2}(r) 
	\nabla_\perp Y_{lm}] + \sum_{l,m} [V_A(r)({\vec e}_r 
	\times \nabla_\perp Y_{lm})]\ ,
\end{equation}
where the first two terms represent the ``polar part'' (with radial
coefficients $V_{P1}$ and $V_{P2}$), while the third one represents the
``axial part'' (with radial coefficient $V_{A}$). Here, ${\vec e}_r$ is
the radial unit vector and, as a result, the axial terms in
(\ref{dec_vh}) have only components in the ${\vec e}_{\theta}$ and ${\vec
e}_{\phi}$ directions.

	A considerable simplification in the treatment of nonradial
oscillations in Newtonian nonrotating fluid stars comes from the fact
that no axial term appears in Euler equations to produce an axial
restoring force. Stated differently, if we decomposed ${\vec \xi}$ as
\begin{equation}
\label{gen_xi}
{\vec \xi} = \xi {\vec e}_r + {\vec \xi}_{\bot} = 
	e^{-i\sigma t} \left[ \xi(r) Y_{lm} {\vec e}_r
	+ \xi^{\rm ax}(r) \nabla_{\bot} Y_{lm} \right ]\ ,
\end{equation}
then all of the terms of equations
(\ref{nr_perts_21})--(\ref{nr_perts_23}) involving ${\vec \xi}_{\bot}$
can be eliminated using the condition [cf. eq. (\ref{nr_perts_22})]
\begin{equation}
-\sigma^2{\vec \xi}_{\bot}
	+ \nabla_{\bot}\left (\frac{\delta p}{\rho_0} + 
	\delta\Phi \right) = 0
	\ .
\end{equation}
As a result, only the polar parts need to be considered. (Note that this
is not the case for stars with a solid crust, or with rotation, or with a
magnetic field.). 

	Assuming now the usual harmonic dependence in time, $\delta p,
\delta \rho, \delta \Phi, \xi \sim e^{-i\sigma t}$, and noting that
equations with different quantum numbers $l,m$ are decoupled (so that the
sums can be replaced by a single pair of $l,m$ indices) one obtains
\begin{eqnarray}
\label{Eqa1}
&& \frac{1}{r^{2}}\frac{d}{dr}\left( r^{2}\xi\right)
	-\frac{g}{c^{2}_{s}}\xi+\left( 1-\frac{L^{2}_{l}}{\sigma^{2}}\right)
	\frac{\delta p}{\rho c^{2}_{s}}=
	\frac{l(l+1)}{\sigma^{2}r^{2}}\delta \Phi \ ,
\\ \nonumber \\
\label{Eqa2}
&& \frac{1}{\rho }\frac{d}{dr}\delta p+\frac{g}{\rho c^{2}_{s}}
	\delta p +\left( N^{2}-\sigma^{2}\right) \xi
	=-\frac{d}{dr}\delta \Phi \ ,
\\ \nonumber \\
\label{Eqa3}
&& \frac{1}{r^{2}}\frac{d}{dr}\left( r^{2}\frac{d}{dr}\delta \Phi \right)
	-\frac{l(l+1)}{r^{2}}\delta \Phi =
	4\pi G\rho \left( \frac{\delta p}
	{\rho c^{2}_{s}}+\frac{N^{2}}{g}\xi\right) \ ,
\end{eqnarray}
where $c^{2}_{s}$ is the local sound speed 
\begin{equation}
c_{s} \equiv \sqrt{\frac{\Gamma p}{\rho }} \ ,
\end{equation}
and $L^{2}_{l}$ is the square of the {\it Lamb frequency} for the $l$-th
mode
\begin{equation}
\label{l_f}
L^{2}_{l}\equiv \frac{l(l+1)}{r^{2}}c^{2}_{s} \ ,
\end{equation}
or, equivalently, $L^{-1}_l$ represents the typical crossing time of
sound waves of wavelength $\lambda_h \simeq 2\pi r/l$ on a $r= {\rm
const.}$ surface.

	The set of equations (\ref{Eqa1})--(\ref{Eqa3}) fully describes
the nonradial oscillation modes of a nonrotating Newtonian star and its
solution requires, in general, a numerical calculation. I will not
discuss here the details of how to obtain such a numerical solution but
will concentrate, in the following Section, on using a simplified form of
equations (\ref{Eqa1})--(\ref{Eqa3}) to derive a first classification of
the different oscillation modes that are solutions of this set of
equations.

\subsection{Classification of Stellar Oscillation Modes}
\label{cosom}

	The initial step in the description of a complex phenomenology
always consists of a ``classification'' and stellar oscillation modes are
no different in this respect. In the case of nonradial modes of
nonrotating stellar models, the classification is particularly simple
when the Eulerian perturbations of gravitational potential can be
neglected (i.e. $\delta \Phi = 0$), so that the perturbed Poisson
equation (\ref{Eqa3}) is trivially satisfied. This is usually referred to
as the {\it Cowling approximation}~\cite{c41} and for a fluid
configuration which is not self-gravitating (i.e. one whose gravitational
potential is such that $\Phi_0 + \delta \Phi = 0$), the Cowling
approximation is actually an exact description of the
pulsations~\cite{il92}. The error made in employing the Cowling
approximation depends rather sensitively on the specific mode
investigated and for modes with predominant axial nature, this error can
be rather small (i.e. $< 10\%$), while it can become more significant
with predominantly polar modes~\cite{ye95,ye97}.

	As customary in the analysis of local oscillations, I now
introduce two new variables $\chi$ and $\eta$, replacing $\delta p$ and
$\xi$, and defined as
\begin{eqnarray}
&&\chi \equiv r^{2}\xi \exp \left(\int_0^r
	\frac{g}{c^{2}_{s}}dr \right) \ ,
\\ \nonumber \\
&&\eta \equiv \frac{\delta p}{\rho }\exp \left( -\int_0^r
	\frac{N^{2}}{g}dr\right)\ .
\end{eqnarray}
As a result, equations (\ref{Eqa1}) and (\ref{Eqa2}) can be rewritten in
the matrix form
\begin{equation}
\frac{d}{dr}\left( \begin{array}{c}
\chi \\
\eta 
\end{array}\right) =\left( 
	\begin{array}{cc}
	0 & \displaystyle \frac{h r^{2}}{c^{2}_{s}}\left( 
	\frac{L^{2}_{l}}{\sigma^{2}}-1\right) 
\\ \nonumber \\
\displaystyle \frac{\sigma^{2}-N^{2}}{r^{2}h} & 0
	\end{array}\right) 
	\left( \begin{array}{c}
	\chi \\
	\eta 
\end{array}\right) \ ,
\end{equation}
where
\begin{equation}
h(r) \equiv \exp \left[ \int_0^r \left( \frac{N^{2}}{g} - 
	\frac{g}{c^{2}_{s}}\right) dr\right] \ .
\label{vectform}
\end{equation}

\noindent Next, I consider the perturbations to have a harmonic spatial
dependence in the radial direction expressed as
\begin{equation}
\label{harmonic}
\chi, \eta \sim e^{ik_r r} \ , 
\end{equation}
where $k_r$ is the radial wavenumber, assumed to be larger than the
length-scale of radial variations in the equilibrium configuration (this
is also known as the WKB approximation and is a distinctive feature of a
``local approach''). Introducing this ansatz in equation
(\ref{vectform}), one obtains the following dispersion relation for
nonradial oscillations in nonrotating stars
\begin{equation}
\label{nr_dr}
k^{2}_r=\frac{1}{\sigma^{2}c^{2}_{s}}\left( \sigma^{2}-L^{2}_{l}\right) 
	\left( \sigma^{2}-N^{2}\right) \ .
\end{equation}

	If $k^{2}_r > 0$, the wavenumber $k_r$ is real and two travelling
waves will exist. The physical interpretation of the dispersion relation
(\ref{nr_dr}) is simpler in the two limiting cases in which either $N=0$
or $L_l=0$, so that the dispersion relation assumes the simpler forms
\begin{equation}
\label{waves_1}
\sigma^{2} = L^{2}_{l} + k^{2}_r c^{2}_{s} \ , 
\end{equation}
and 
\begin{equation}
\label{waves_2}
\sigma^{2} = N^{2} + k^{2}_r c^{2}_{s} \ . 
\end{equation}
Furthermore, in the long wavelength limit $k_r =0$, the two frequencies
in (\ref{waves_1})--(\ref{waves_2}) correspond respectively to the Lamb
frequency $\sigma = L_l$ [cf. eq. (\ref{l_f})] and to the
Brunt-V\"{a}is\"{a}l\"{a} frequency $\sigma = N$ [cf. eq. (\ref{bv_f})].

	If, on the other hand, $k^{2}_r < 0$, the wavenumber $k_r$
is imaginary, corresponding to waves that are exponentially damped
somewhere in the star and are thus said to be ``evanescent''.

\begin{figure}[htb]
\begin{center}
\hspace{0.125truecm}
\leavevmode
\hbox{\psfig{figure=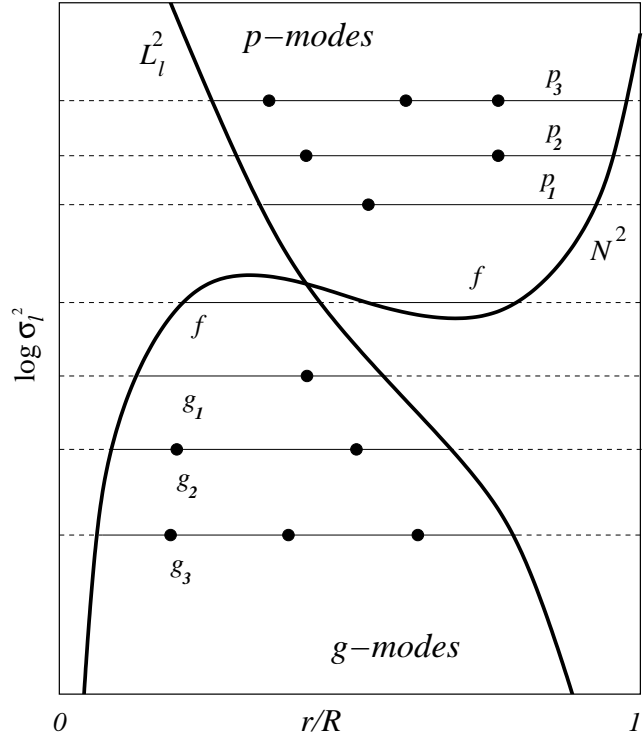,width=16.0truecm,angle=-90} }
\caption{\footnotesize Schematic propagation diagram in a simple stellar
model. The two thick lines represent the Lamb and
Brunt-V\"{a}is\"{a}l\"{a} frequencies respectively, while the horizontal
lines show the location of eigenfrequencies of several modes. Dashed
lines are used to show the position within the star at which the waves
become evanescent, while solid lines refer to eigenmode solutions; filled
dots on these lines show the radial location of the nodes of the
corresponding eigenfunctions.
\label{fig5}}
\end{center}
\end{figure}

	All of this is summarized in Figure~\ref{fig5} which shows the
frequencies of the different modes of oscillations as a function of the
radial position within the star. Such a diagram is also referred to as
the {\it propagation diagram} of the dispersion relation (\ref{nr_dr})
for a given number $l$. The thick solid lines represent the values of the
to the Lamb and Brunt-V\"{a}is\"{a}l\"{a} frequencies in the long
wavelength limit $k_r =0$ and divide the diagram in four different
regions. In two of these regions the waves are evanescent and the
corresponding frequencies are indicated with dashed lines. In the
remaining two regions, (referred to as the upper and lower branch,
respectively), the wavenumbers are real and the corresponding waves can
be trapped to form an eigenmode (or global) oscillation. The frequencies
of such modes are indicated with solid lines and the number of filled
dots indicates the position of the nodes in the corresponding
eigenfunctions. In what follows I will use Figure~\ref{fig5} to provide a
brief description of the different modes that can be classified in an
ordinary star, distinguishing those with pure polar properties from those
with pure axial ones.

\subsubsection{Even-parity (Polar) Modes}
	
\begin{itemize}

\item{\it $p$-modes -- } They fill the high-frequency branch of the
propagation diagram and have pressure gradients as the main restoring
force. The eigenfunctions of a mode with quantum number $n$ has $n$
nodes. As $n$ increases, the frequency increases and the wavelength
becomes smaller.  In the limit of short wavelength, these modes
represents simple acoustic waves travelling in the star. In this same
limit the frequency of the mode will tend to infinity.

	For a ``canonical'' neutron star, i.e. a neutron star with mass
$M=1.4 M_{\odot}$ and radius $R\simeq 14$ km, the typical frequency of
the lowest order $p$-mode is a few kHz.

\item{\it $g$-modes -- } They fill the low-frequency branch of the
propagation diagram and have a buoyancy force (produced, for instance, by
gradients in temperature, composition or density) as the main restoring
force. The eigenfunctions of a mode with quantum number $n$ has $n$
nodes. As $n$ increases, the frequency decreases and the wavelength
becomes smaller. In the limit of short wavelength the
frequency of the mode goes to zero.  

	In neutron stars, two different types of $g$-modes are
expected. One of them can be of modes trapped in the high density core
region of the star and has a typical frequency of $\sim 0.1$ kHz. The
other $g$-modes are expected to be trapped in the outer layers of the
star (i.e. in the fluid ``ocean'' above the crust) and have a typical
frequency of $\sim $ 10 Hz.

\item{\it $f$-modes -- } These modes have a character which is
intermediate between those of $p$- and $g$-modes and are also referred to
as the {\it fundamental} modes of oscillation. For each $l$, the
frequency of this mode is between the lowest order $g$-mode (i.e. the
highest frequency $g$-mode) and the lowest order $p$-mode (i.e. the
lowest frequency $p$-mode). Note that for a given pair of quantum numbers
$(l, m)$, only one $f$-mode exists and its eigenfunctions have no
nodes. For an incompressible homogeneous sphere, the $f$-mode represents
the only oscillation mode and its frequency scales with the mean density
\begin{equation}
\label{f_freq_inc}
\sigma_{f,l} = \sqrt{\frac{2l(l+1)}{2l+1}\left(\frac{M}{R^3}\right)} \ .
\end{equation}
	The typical frequency of the lowest order $f$-mode for a
canonical nonrotating neutron star is $\sim 2-3$ kHz with a decaying time
of $\sim 0.1-0.5$ s.

\end{itemize}

\subsubsection{Odd-parity (Axial) Modes}

	As mentioned in Section~\ref{nr_ns}, this type of modes has zero
frequency for nonrotating stars that do not have a crust or a magnetic
field. When rotation is present, however, these modes are no longer
degenerate and give rise to a rich and complex phenomenology. 

\begin{itemize}
\item{\it inertial modes (including r-modes) -- } These modes have the
Coriolis force as the main restoring force. In Newtonian stars, inertial
modes have velocity fields described by a mixture of polar- and
axial-parity components to first order in the slow-rotation
expansion. The so-called ``$r$-modes'' are a special class of inertial
modes and have axial-parity components only. The velocity eigenfunctions
have very simple expressions and the perturbations in the density and
pressure appear at orders higher than the first one in the slow-rotation
approximation. The Newtonian $r$-modes are indeed a generalization of the
Rossby waves, well-known in the geophysical hydrodynamics and follow a
simple dispersion relation that, in a frame corotating with the star,
takes the form
\begin{equation}
\sigma_r =  - \frac{2 m \Omega_i}{l(l+1)} \ ,
\end{equation}
where $\Omega_i$ is the rotational angular frequency of the star as
measured in an inertial frame. Since $\sigma_r \Omega_i <0$, the wave
pattern is retrograde in the rotating frame and in an inertial frame
(i.e. in the reference frame of a distant observer) the mode will be
observed to have frequency
\begin{equation}
\label{in_vs_rot}
\sigma_i = m \Omega_i - \sigma_r = \frac{(l-1)(l+2)}{l+1} \Omega_i 
	\ .
\end{equation}

	A good starting point in the analysis of the classical $r$-modes
in Newtonian stars can be found in~\cite{s82}, while a review of the
impact of $r$-modes in the development of an instability can be found
in~\cite{ak01}. Finally, a discussion of the interaction of $r$-modes
with the magnetic fields that are likely to be present in a neutron star
can be found in~\cite{rls00,rlms01}.

\end{itemize}

\subsubsection{Purely Relativistic Modes}

	As discussed in Section~\ref{intro_gws}, a general relativistic
spacetime has its own dynamical degrees of freedom, thus allowing for the
existence and propagation of gravitational waves. This distinctive
feature of General Relativity gives rise to stellar oscillation modes
that do not have a counterpart in Newtonian stars. These modes are called
$w$-modes and were first shown to exist only rather
recently~\cite{ks92}. These modes are characterized by spacetime
fluctuations that couple with the matter only very weakly. As a result,
the stellar fluid is left almost unperturbed during $w$-mode
oscillations. These purely relativistic modes can be further classified
according to their properties which I briefly summarise below. More
information can be found in the review work by Kokkotas and
Schmidt~\cite{ks99}.

\begin{itemize}

\item {\it Curvature Modes}: they are the standard $w$-modes~\cite{ks92}
and are reminiscent of the QNMs in black holes. The main features of
these modes is that they have rather high frequencies of $\sim 5-12$ kHz
and very rapidly decaying times of $\sim 0.02-0.1$ ms, with the damping
rate increasing as the compactness of the star decreases. Modes of higher
order have higher frequencies and shorter damping times.

\item {\it Trapped Modes}: they exist only for supercompact stars ($R \le
3M$). Their existence is due to the fact that for such compact stars, the
stellar surface is inside the peak of the gravitational field's potential
barrier~\cite{cf91} (cf. Fig.~\ref{fig3}). The trapped modes are finite
in number (i.e. $\sim 7$) and their number increases as the potential
well becomes deeper (i.e. as the star becomes more and more
compact). Trapped modes differ from curvature modes in that their damping
times are considerably longer (i.e. a few tenths of a second), with
frequencies that are from a few hundred Hz to a few kHz. At present it is
not yet clear whether such ultra-compact stars can be built through a
realistic equation of state.

\item {\it Interface Modes}: they resemble acoustic waves scattered off a
hard sphere but do not induce significant fluid motion~\cite{leins93}.
For typical neutron stars they have frequencies in the range $2-15$ kHz
and are extremely rapidly damped, with damping times are of the order of
less than a tenth of a millisecond.

\end{itemize}
More on $w$-modes will be discussed in the following Section.

\subsubsection{Some Empirical Expressions for ${\boldsymbol f\,}$-,
${\boldsymbol p\,}$- and ${\boldsymbol w\,}$-modes }

	As mentioned in the previous Sections, the different modes of
oscillation depend on the physical properties of the stellar model and
can provide detailed information on them.  Nevertheless, the frequencies
of the lowest order modes are not very sensitive on the fine details of
the stellar structure and can be estimated through simple empirical
formulas. This approach was first suggested by Andersson and
Kokkotas~\cite{ak98} and has lead to the different empirical expressions
I will discuss below.

	Since the relation between the square of the $f$-mode frequencies
and the mean density is linear [cf. equation (\ref{f_freq_inc})],
introducing the following normalized quantities: ${\bar M} \equiv M /
({1.4 M_\odot})$ and ${\bar R} \equiv R/(10\ {\rm km})$, allows one to
express the $f$-mode frequencies through the simple relation

\begin{equation}
\nu_f \approx 0.78 + 1.63 \left( \frac{\bar M}{{\bar R}^3} 
\right)^{1/2} \ \ {\rm kHz} \ ,
\label{rfw}
\end{equation}
which shows that the typical $f$-mode frequency is expected to be around
2.4 kHz. Equally important is to deduce a corresponding relation for the
damping rate of the $f$-mode and this can be roughly estimated as
\begin{equation}
\tau_f \sim \frac{\mbox{(oscillation energy)}} 
	{\mbox{(power emitted in gravitational waves)}}
	\sim R\left(\frac{R}{M}\right)^3 \ ,
\end{equation}
where the power emitted in gravitational waves can be deduced from the
quadrupole formula. Doing so leads to a relation of the type
\begin{equation}
\frac{1}{\tau_f} \approx \frac{{\bar M}^3} {{\bar R}^4}
\left[ 22.85 - 14.65 \left( \frac{{\bar M}} {\bar R} \right)\right] 
	\ \ {\rm s}^{-1} \ ,
\label{rftau}
\end{equation}
which indicates that a typical $f$-mode oscillation is damped over about
0.6 s.

	A similar procedure can be followed for the oscillation frequency
of $p$-modes, which can be roughly estimated to scale as 
\begin{equation}
\nu_p \approx \frac{1}{{\bar M}}\left(1.75 + 5.59 
	\frac{\bar M}{\bar R} \right) \ \ {\rm kHz} \ ,
\label{rpw}
\end{equation}
so that the typical $p$-mode frequency is around 7 kHz. Unfortunately,
the damping of $p$-modes is sensitive to changes in the modal
distribution inside the star and different EOSs lead to rather different
$p$-mode damping rates. As a result, no robust empirical expression can
be derived in this case. 

	This is clearly not the case for $w$-modes that do not excite a
significant fluid motion and are therefore only weakly dependent on the
characteristics of the fluid. Rather, it can be shown analytically that
the frequency of $w$-modes is inversely proportional to the size of the
star~\cite{ks92}, while the damping time is a function of the
compactness~\cite{akk96}.  As a result, rather robust relations can be
found for the frequency of the first $w$-mode
\begin{equation}
\nu_w \approx \frac{1}{\bar R}
        \left[ 20.92 - 9.14 \left( \frac{\bar M}{\bar R} \right)
	\right] \ \ {\rm kHz} \ ,
\label{rww}
\end{equation}
and for its damping time 
\begin{equation}
\frac{1}{\tau_w} \approx \frac{1}{\bar M}
	\left[ 5.74 + 103 \left( \frac{\bar M}{\bar R} \right)
	- 67.45 \left( \frac{\bar M}{\bar R} \right)^2 \right] 
	\ \ {\rm ms}^{-1} \ .
\label{rwtau}
\end{equation}
Expressions (\ref{rww}) and (\ref{rwtau}) show that $w$-modes have
typical frequencies around 12 kHz (but these can vary sensibly with the
radius of the star) and a damping time of $\sim 0.02$ ms (which is
comparable to that of an oscillating black hole with the same mass).

\subsection{Non-axisymmetric Instabilities}

 	Some of the non-axisymmetric modes of oscillation in rotating
stars may not be damped, but have amplitudes that grow exponentially in
time. When this is the case, the oscillations are said to be "unstable"
and the resulting instability can either be {\it dynamical}, if it
develops on the timescale set by the rotation or by the free-fall, or
{\it secular}, if it develops on a much longer timescale set, for
instance, by dissipative processes. Dynamical instabilities differ
considerably from secular ones in that they are purely hydrodynamical,
while the latter are triggered by dissipative process such as viscous
dissipation, emission of gravitational or electromagnetic radiation,
thermal losses, etc.. In both cases, however, the instabilities reflect
the attempt of the rotating star to find a lower energy state either by
changing its mass distribution (e.g. through variations of the momentum
of inertia) or by violating the conservation of some quantity
(e.g. circulation or angular momentum). A quantity which is often used to
measure how close the rotating star is to the onset of an instability is
the so called {\it rotational parameter} $\beta$
\begin{equation}
\label{beta}
\beta \equiv \frac{\mbox{(rotational kinetic energy)}}
	{\mbox{(gravitational energy)}} \equiv \frac{T}{|W|} 
	\approx \frac{1}{3}
	\left(\frac{\Omega}{\Omega_{_{\rm K}}}\right)^2
	\ ,
\end{equation}
The last equality in (\ref{beta}) has been derived for a Newtonian star,
with $\Omega/\Omega_{_{\rm K}}$ being the stellar angular velocity
normalized to the Keplerian value, that is, the value of the angular
velocity at which matter can be shed at the stellar equator. Indicating
with ${\bar \rho}$ the average rest-mass density, the Keplerian angular
velocity can be estimated to be $\Omega_{_{\rm K}} \sim (2/3) \sqrt{\pi
{\bar \rho}}$. 

	The parametrization (\ref{beta}) is independent of the rotation
law and is particularly useful for differentially rotating objects. By
definition and invoking the Virial theorem, the parametrization is
constrained to be between $\beta=0$ (for a spherical object) and
$\beta=1/2$ (for an infinitely extended, thin disc at rest)~\cite{t78}. A
well-known application of the rotational parameter (\ref{beta}) is
offered by classical result for the onset of the dynamical instability in
Newtonian rotating stars, and which has been estimated to be $T/|W|
\simeq 0.27$ for a variety of different equations of state and rotation
laws.

	Since the secular non-axisymmetric instabilities are triggered by
dissipative mechanisms, their development will be different according to
whether they are driven by viscous processes or by the emission of
radiation (either gravitational or electromagnetic). When viscous
dissipation processes are present and the radiative losses are
negligible, an initially axisymmetric, incompressible rotating object,
i.e. a {\it Maclaurin spheroid}, will be deformed into a {\it Jacobi
ellipsoid}, i.e. into a uniformly rotating, homogeneous configuration
with ellipsoidal surfaces (they are similar to rotating american
``footballs''). This happens at roughly $T/|W| \simeq 0.14$ and is
referred to as the viscous-driven $f$-mode instability (see~\cite{st84}
for a complete discussion).

	When viscous processes are negligible, on the other hand, the
growth of the non-axisym-\linebreak metric modes can be driven by the
emission of gravitational or electromagnetic radiation (although the
latter is usually much smaller than the former). The instability that
develops in this way is the so called CFS (Chandrasekhar-Friedman-Schutz)
instability~\cite{c70,fs78} and is produced by the coupling between the
loss of energy and angular momentum via radiation and the
non-axisymmetric oscillations modified by the stellar rotation (A
qualitative introduction to the CFS instability is presented in Appendix
A). Also in this case, an initially axisymmetric Maclaurin spheroid will
be deformed into a uniformly rotating, homogeneous configuration with
ellipsoidal surfaces, called {\it Dedekind ellipsoid}. The difference
between the Jacobi and Dedekind ellipsoids is that in the latter the
ellipsoidal surfaces are supported by internal circulations but the shape
is stationary as observed by an inertial observer (they are therefore
similar to nonrotating american ``footballs''). This happens again at
roughly $T/|W| \simeq 0.14$ and is referred to as the $f$-mode
CFS-instability.

	In practice, the development of non-axisymmetric instabilities is
much more complicated than what discussed in the two limiting cases
above, because {\it both} viscous and radiative losses are active at the
same time in realistic stars. As a result, the modes that are driven
unstable by viscosity and deform a Maclaurin spheroid into a Jacobi
ellipsoid (i.e. the Jacobi modes) tend to be stabilized by the emission
of gravitational waves (the star develops non-axisymmetric ``ripples'' to
remove the excessive angular momentum via the emission of gravitational
waves). At the same time, the modes that are driven unstable by the
emission of gravitational waves and deform a Maclaurin spheroid into a
Dedekind ellipsoid (i.e. the ``Dedekind modes'') tend to be stabilized by
the viscous dissipative processes (the increased shear stresses, for
example, tend to remove, with the aid of the shear viscosity, the
non-axisymmetric ``ripples'' responsible for the emission of
gravitational waves). Computing the delicate balance between these two
mechanisms in regimes of strong gravitational fields, high-density matter
and temperatures is extremely difficult and at the core of the present
research on the emission of gravitational waves from instabilities in
relativistic stars.

       Because of the very large amplitudes that the oscillation modes
can reach when driven unstable, the amount of gravitational radiation
emitted can become considerable and these unstable stars can then become
promising sources of gravitational waves, potentially detectable by the
gravitational-wave observatories now working or being under construction.
A recent review on this fascinating area of research can be found
in~\cite{a02}.

\section*{Acknowledgments}

	It is a pleasure to thank the organizers of the School and in
particular Goran Senjanovic and Francesco Vissani for their assistance in
organizing these lectures. I am also indebted with Shin'ichirou Yoshida
for his valuable help in writing the Section on stellar perturbations and
with Valeria Ferrari for the numerous discussions and for carefully
reading these notes. Finally, special thanks go to the SISSA 2001-2002
``first-year'' graduate students for their help in typing some of my
lecture notes of the General Relativity course.  Financial support for
this research has been provided by the MIUR and by the EU Network
Programme (Research Training Network Contract HPRN-CT-2000-00137).


\newpage
\appendix

\section{A Qualitative Introduction to the CFS Instability}
\label{cfsi}
\def\theequation{\thesection.\arabic{equation}}
\setcounter{equation}{0}

	The instability was first discovered by Chandrasekhar~\cite{c70},
and subsequently considered by Friedman and Schutz~\cite{fs78}, who have
shown its generic nature. While a formal proof of the criteria for the
instability are rather involved~\cite{fs78}, qualitative arguments on the
properties of the instability can be given simply using a couple of
illustrative examples.

	Consider, therefore, a rotating star which is undergoing
non-axisymmetric oscillations. For simplicity I will consider the
simplest non-axisymmetric perturbation with mode numbers $l=m=2$. Because
the star is rotating, the properties of the perturbations can be
considered both in a reference frame which is corotating with it
(i.e. the ``rotating'' frame) or in a reference frame which is not
rotating and is fixed with respect to, say, distant stars (i.e. the
``inertial'' frame).

\begin{figure}[htb]
\begin{center}
\hspace{0.125truecm} \leavevmode
\hbox{\psfig{figure=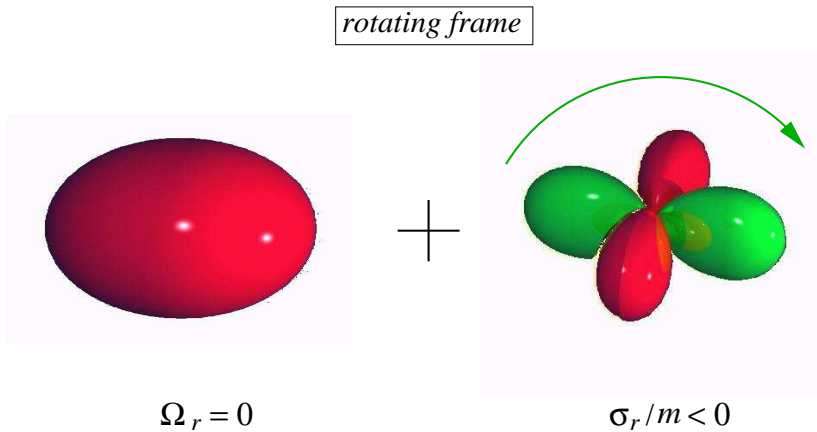,width=12.0truecm,angle=-90} }
\caption{\footnotesize Schematic view from the rotating frame. The
unperturbed star is shown on the left and the non-axisymmetric
perturbation on the right. $\Omega_r=0$ and $\sigma_r/m$ are the angular
velocities of the star and of the wave-pattern, respectively. Note that
the star is effectively rotating as observed in an inertial frame and its
structure, therefore, is deformed into a spheroid by the centrifugal
stresses.
\label{fig7}}
\end{center}
\end{figure}

	I will first discuss what would be observed in the rotating
frame; this is summarized in Fig.~\ref{fig7} where I have shown
schematically the unperturbed star on the left and the non-axisymmetric
$l=m=2$ perturbation on the right. In the corotating frame the star has a
zero angular velocity (i.e. $\Omega_{r}=0$), but is nevertheless deformed
into a spheroid by the centrifugal force. The perturbation on the other
hand, has a nonzero frequency $\sigma_r$ and the corresponding $m=2$
wave-pattern is seen to rotate with angular frequency $\sigma_r/m$ which
is, say, negative. By definition, such a mode is referred to as {\it
``retrograde''} and because the perturbed star is rotating at an angular
velocity smaller than the initial one, the mode has a {\it negative}
angular momentum $J_0<0$ in the corotating frame.

	The non-axisymmetric perturbation generates a time-variation of
the stellar mass multipoles (and/or of the mass-current multipoles) and
the gravitational waves that are produced in this way, carry positive
amounts energy at infinity. The angular momentum carried at infinity
$j_{\rm gw}$ , on the other hand, can either be positive or negative
according to the sense in which the perturbation is seen to rotate in the
inertial frame. I therefore need to consider how the perturbation is
observed in the inertial frame which, I recall, is the frame in which
quantities like the total amount of energy and angular momentum can be
measured unambiguously. 

\begin{figure}[htb]
\begin{center}
\hspace{0.125truecm}
\leavevmode
\hbox{\psfig{figure=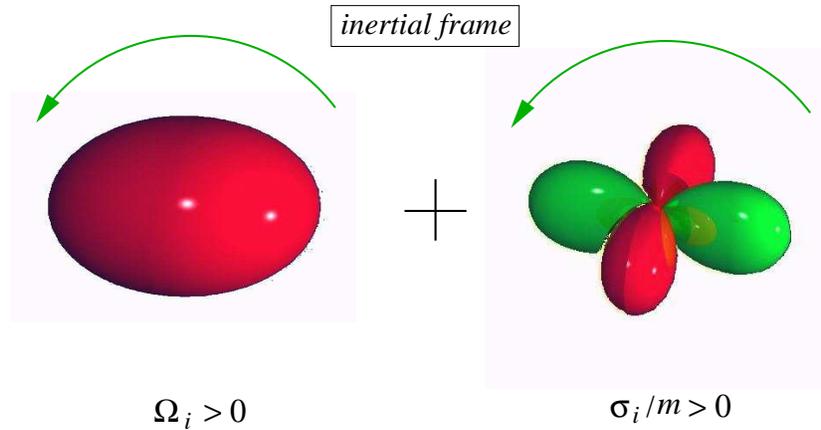,width=12.0truecm,angle=-90} }
\caption{\footnotesize Schematic view from the inertial frame. The
unperturbed star is shown on the left and the non-axisymmetric
perturbation on the right. $\Omega_i$ and $\sigma_i/m$ are the angular
velocities of the star and of the wave-pattern, respectively. Note that
the star is seen to rotate and appears therefore as a spheroid.
\label{fig8}}
\end{center}
\end{figure}

	The ``view'' from the inertial frame is summarised in
Fig.~\ref{fig8} where, again, I have shown schematically the unperturbed
star on the left and the non-axisymmetric perturbation on the right. An
observer in this frame will then see the star rotating at a nonzero, say
positive, angular velocity $\Omega_i>0$ and the non-axisymmetric
perturbation with a wave pattern that is also rotating with angular
velocity $\sigma_i/m$.

	It is not difficult to realize that the direction in which the
wave-pattern rotates depends on both $\Omega_i$ and $\sigma_r$ through
the relation between the frequencies in the two reference frames:
$\sigma_i = m \Omega_i - \sigma_r$ [cf. expression (\ref{in_vs_rot})]. As
a result, the wave-pattern can either be ``dragged'' forward ($\sigma_i >
0 >\sigma_r $), or backward ($\sigma_i < \sigma_r < 0$) by the stellar
rotation $\Omega_i$. The obviously interesting case shown in
Fig.~\ref{fig8}, is the one in which the mode is dragged forward
(i.e. $\sigma_i/m>0$) and the non-axisymmetric mode, which is then said
to be {\it ``prograde''}, is seen to rotate in the same sense as the
rotating star (Note that the ``dragging'' of the wave-pattern is a purely
kinematical and Newtonian effect, fundamentally distinct from the general
relativistic ``dragging of reference frames''.). When this happens,
$\Omega_i \sigma_r < 0$ and the conditions for the onset of the CFS
instability are met. In this case, in fact, and because the sign of the
angular momentum lost is determined by the sense of rotation of the
oscillation's wave-pattern, the prograde non-axisymmetric perturbation
will carry to infinity {\it positive} amounts of angular momentum,
i.e. $j_{\rm gw} > 0$.

	For the observer in the rotating frame, on the other hand, the
total angular momentum of the mode $J(t) \equiv J_0 - j_{\rm gw}(t)$
becomes increasingly negative because of the losses through $j_{\rm
gw}(t)$ that continuously reduce $J(t)$, i.e. $J(t)<J_0<0$.  As a result,
the initially small non-axisymmetric perturbation with negative angular
momentum in the corotating frame, is driven to large amplitude
oscillations with a progressively larger negative angular momentum. Such
perturbation emits increasingly large amounts of gravitational waves,
thus feeding the development of the instability. Using a pictorial
analogue, the development of the CFS is similar to someone's debts that
get larger as new expenses (with positive amounts of money) are made.
The growth of the instability stops when either nonlinear or dissipative
effects become important and transfer energy from the unstable mode into
the other available channels.



\end{document}